\documentclass[aps,prd,nofootinbib,onecolumn,superscriptaddress,preprintnumbers,balancelastpage,longbibliography,nobibnotes]{revtex4-1}
\usepackage{graphicx,epsfig,psfrag,bm,amssymb}
\usepackage{dcolumn}
\usepackage{bm}
\usepackage[dvipsnames]{xcolor}
\usepackage{braket}
\usepackage{mathrsfs,amsfonts,hepunits,color}
\usepackage{hyperref}
\usepackage{comment}
\usepackage{soul}

\definecolor{linkcolor}{rgb}{0.0, 0.28, 0.67}
\hypersetup{colorlinks=true,
linkcolor=black,
citecolor=linkcolor,
urlcolor=linkcolor,
linktocpage=true
}

\DeclareRobustCommand{\Eq}[1]{Eq.~(\ref{#1})}
\DeclareRobustCommand{\Eqs}[2]{Eqs.~(\ref{#1}) and (\ref{#2})}
\newcommand{\Sec}[1]{Sec.~\ref{sec:#1}}
\newcommand{\App}[1]{Appendix~\ref{app:#1}}
\newcommand{\Fig}[1]{Fig.~\ref{fig:#1}}

\newcommand{\be}{\begin{equation}}
\newcommand{\ee}{\end{equation}}
\newcommand{\E}{\mathbf{E}}
\newcommand{\B}{\mathbf{B}}

\newcommand{\n}{\mathbf{n}}

\newcommand{\mn}{\mu \nu}
\newcommand{\rs}{\rho \sigma}
\newcommand{\hmn}{h_{\mu \nu}}
\newcommand{\TT}{\text{TT}}
\newcommand{\jeff}{j_\text{eff}}
\newcommand{\hnorm}{h_0}
\newcommand{\order}[1]{\mathcal{O}{(#1)}}
\newcommand{\nl}{\nonumber \\}
\newcommand{\w}{\omega}
\newcommand{\wg}{\omega_g}

\newcommand{\tTT}{t_{_\text{TT}}}
\newcommand{\xTT}{x_{_\text{TT}}}
\newcommand{\yTT}{y_{_\text{TT}}}
\newcommand{\zTT}{z_{_\text{TT}}}
\newcommand{\UTT}{U_{_\text{TT}}}

\newcommand{\FTT}{F_{_\text{TT}}}

\newcommand{\jv}{\boldsymbol{j}}
\newcommand{\jveff}{\boldsymbol{j}_\text{eff}}
\newcommand{\jvhat}{\hat{\jv \, }}
\newcommand{\jplus}{\jvhat_{\hspace{-0.1 cm} +}}
\newcommand{\jcross}{\jvhat_{\hspace{-0.1 cm} \times}}
\newcommand{\jpc}{\jvhat_{\hspace{-0.1 cm} +, \times}}

\newcommand{\xv}{{\bf x}}
\newcommand{\Vcav}{V_\text{cav}}

\begin{document}

\preprint{FERMILAB-PUB-21-724-SQMS-T}

\title{Detecting High-Frequency Gravitational Waves with Microwave Cavities}

\author{Asher Berlin}
\affiliation{Center for Cosmology and Particle Physics, Department of Physics,
New York University, New York, NY 10003, USA}
\affiliation{Theoretical Physics Division, Fermi National Accelerator Laboratory, Batavia, IL 60510, USA}
\affiliation{Superconducting Quantum Materials and Systems Center (SQMS), Fermi National Accelerator Laboratory, Batavia, IL 60510, USA}
\author{Diego Blas}
\affiliation{Grup de F\'{i}sica Te\`{o}rica, Departament de F\'{i}sica, Universitat Aut\`{o}noma de Barcelona, 08193 Bellaterra, Spain}
\affiliation{Institut de Fisica d’Altes Energies (IFAE), The Barcelona Institute of Science and Technology, Campus UAB, 08193 Bellaterra (Barcelona), Spain}
\author{Raffaele Tito D’Agnolo}
\affiliation{Universit\'e Paris-Saclay, CEA, Institut de Physique Th\'eorique, 91191, Gif-sur-Yvette, France}
\author{Sebastian~A.~R.~Ellis}
\affiliation{D\'epartement de Physique Th\'eorique, Universit\'e de Gen\`eve, 
24 quai Ernest Ansermet, 1211 Gen\`eve 4, Switzerland}
\affiliation{Universit\'e Paris-Saclay, CEA, Institut de Physique Th\'eorique, 91191, Gif-sur-Yvette, France}
\author{Roni Harnik}
\affiliation{Theoretical Physics Division, Fermi National Accelerator Laboratory, Batavia, IL 60510, USA}
\affiliation{Superconducting Quantum Materials and Systems Center (SQMS), Fermi National Accelerator Laboratory, Batavia, IL 60510, USA}
\author{Yonatan Kahn}
\affiliation{Department of Physics, University of Illinois at Urbana-Champaign, Urbana, IL 61801, USA}
\affiliation{Illinois Center for Advanced Studies of the Universe, University of Illinois at Urbana-Champaign, Urbana, IL 61801, USA}
\affiliation{Superconducting Quantum Materials and Systems Center (SQMS), Fermi National Accelerator Laboratory, Batavia, IL 60510, USA}
\author{Jan Sch\"{u}tte-Engel}
\affiliation{Department of Physics, University of Illinois at Urbana-Champaign, Urbana, IL 61801, USA}
\affiliation{Illinois Center for Advanced Studies of the Universe, University of Illinois at Urbana-Champaign, Urbana, IL 61801, USA}
\affiliation{Superconducting Quantum Materials and Systems Center (SQMS), Fermi National Accelerator Laboratory, Batavia, IL 60510, USA}
\date\today

\begin{abstract}

We give a detailed treatment of electromagnetic signals generated by gravitational waves (GWs) in resonant cavity experiments. Our investigation corrects and builds upon previous studies by carefully accounting for the gauge dependence of relevant quantities. We work in a preferred frame for the laboratory, the proper detector frame, and show how to resum short-wavelength effects to provide analytic results that are exact for GWs of arbitrary wavelength. This formalism allows us to firmly establish that, contrary to previous claims, cavity experiments designed for the detection of axion dark matter only need to reanalyze existing data to search for high-frequency GWs with strains as small as $h\sim 10^{-22}-10^{-21}$. We also argue that directional detection is possible in principle using readout of multiple cavity modes. Further improvements in sensitivity are expected with cutting-edge advances in superconducting cavity technology.
\end{abstract}

\maketitle

\section{Introduction}

The first direct observations of gravitational waves (GWs) by the ground-based interferometers LIGO and Virgo~\cite{LIGOScientific:2016aoc} have ushered in the era of GW astronomy. While the central focus of such experiments has been on the $\text{Hz} - \text{kHz}$ frequency range, an exploration across a much wider spectrum is warranted. The Universe is expected to be populated by GWs over many decades in frequency, analogous to electromagnetic (EM) radiation~\cite{longair_2011}, carrying information that may revolutionize our understanding of Nature. This fact has spurred the development of a large array of observational efforts with the aim of detecting much lower frequency signals compared to current interferometers. These include future ground-based~\cite{Hild:2010id, Punturo:2010zz,LIGOScientific:2016wof} and space-based~\cite{LISA:2017pwj,Yagi:2011wg} laser interferometers, atom interferometers~\cite{Badurina:2019hst,Abe:2021ksx,AEDGE:2019nxb},  pulsar timing arrays~\cite{NANOGrav:2020bcs,Janssen:2014dka}, and CMB observations~\cite{Namikawa:2019tax,CMB-S4:2020lpa}, as well as new types of astrophysical signatures~\cite{Blas:2021mqw}. 

On the other hand, GWs of much higher frequency have more recently garnered renewed interest. A number of interesting proposals and operating detectors for high-frequency GW detection already exist, including interferometers~\cite{Ackley:2020atn, Bailes:2019oma, Akutsu:2008qv, Holometer:2016qoh, Nishizawa:2007tn}, microwave and optical cavities~\cite{Braginskii:1973vm,Grishchuk:1975tg,Braginsky:1979qs,Mensky:2009zz, Caves:1979kq, Pegoraro:1978gv,Pegoraro:1977uv, Reece:1984gv, Reece:1982sc, Ballantini:2005am,Bernard:2002ci,Bernard:2001kp, Ballantini:2003nt, Cruise:2000za, Cruise:2005uq, Cruise:2006zt}, optically levitated sensors~\cite{Aggarwal:2020umq}, mechanical resonators~\cite{Goryachev:2014nna,Goryachev:2014yra, Aguiar:2010kn, Gottardi:2007zn}, superconducting rings~\cite{Anandan:1982is}, and detectors based on the inverse-Gertsenshtein effect~\cite{Ejlli:2019bqj} and the excitation of collective magnon modes~\cite{Ito:2019wcb} (we refer the reader to Ref.~\cite{Aggarwal:2020olq} for a comprehensive review). However, there are still orders of magnitude in both GW frequency and amplitude, well-motivated by theory expectations, that we are currently unable to explore.

In this work, we focus on how GWs couple to electromagnetism, highlighting in particular the role of small-scale laboratory experiments for the detection of GHz-frequency signals. A GW propagating through a static background EM field sources a feeble EM field that oscillates at the frequency of the GW. Resonant detectors are well-suited to the detection of such oscillating fields provided that the GW is coherent over many oscillation cycles. In fact, similar signals arise from other new physics sources, most notably in the case of ultralight axion dark matter that  couples to electromagnetism. Motivated by the tremendous progress in small-scale technology targeting dark matter detection, we focus on setups that are either identical or similar to existing experiments (such as ADMX~\cite{ADMX:2021nhd}, HAYSTAC~\cite{HAYSTAC:2018rwy}, ORGAN~\cite{McAllister:2017lkb}, and CAPP~\cite{CAPP:2020utb}), which feature a resonant conducting cavity of size $L_\text{det} \sim \mathcal{O}({\rm cm}) - \mathcal{O}({\rm m})$ immersed in a strong static magnetic field. Since the resonant frequencies of conducting cavities are comparable to their inverse geometric size, such setups are naturally sensitive to GWs in the GHz regime. Our results also apply to other electromagnetic resonators, such as LC circuits~\cite{Chaudhuri:2014dla}.

From a more general perspective, a second goal of this work is to provide a description of how GWs couple to electromagnetism in a manner that is largely agnostic to the particular experimental setup. In performing such calculations, great care must be taken to preserve gauge invariance (equivalent to consistently incorporating the signal within a particular choice of frame). In particular, GW signals are often computed in the so-called transverse-traceless (TT) gauge, since the spacetime metric is especially simple in this case. However, in this frame, the background EM field and the cavity modes do not coincide with those in flat space. This has not always been taken into account in previous calculations, which has led several studies to conclude that no EM signal is generated when the background magnetic field is aligned with the GW's direction of propagation. As we show in this paper, this statement is at odds with gauge invariance. Our treatment illustrates that existing experiments targeting axions, such as ADMX and HAYSTAC, already have sensitivity to high-frequency GWs and need only to reanalyze existing data with a different signal template. 

For the detectors considered in this work, complications arising from gauge artifacts are avoided by noting that the laboratory defines a preferred frame, the so-called \emph{proper detector} (PD) frame~\cite{Manasse:1963zz,Misner1973,Maggiore}. For this reason, the majority of our calculations adopt the PD frame. However, in order to demonstrate gauge invariance,  we also perform a simple toy example calculation in both the TT and PD frames to show that they yield identical results. More generally, the use of the PD frame has typically been restricted to situations where the GW wavelength $\lambda_g$ is much larger than the size of the detector $L_\text{det}$~\cite{Manasse:1963zz,Misner1973}, such that it suffices to keep only the leading $\order{L_\text{det}^2 / \lambda_g^2}$ corrections to the flat spacetime metric. Here,  we further improve upon such calculations by resumming the GW perturbation to the metric to all orders in $L_\text{det}/\lambda_g$, allowing for the use of the PD frame even when the GW is on resonance with the cavity, which occurs when $L_\text{det} / \lambda_g \sim \order{1}$.

The rest of this paper is organized as follows. In \Sec{general}, we provide a conceptual overview of the class of experimental signals discussed here and derive the GW-EM coupling in the form of an effective current. We also demonstrate gauge invariance between the TT and PD frames with a simple toy example consisting of a GW impinging on a background magnetic field in empty space. This lays the foundations for applying this formalism to a more realistic setup consisting of a resonant cavity immersed in a magnetic field, for which we motivate the optimally-coupled cavity modes in \Sec{currents}. Following a brief survey of possible GW sources in \Sec{sources}, in \Sec{sensitivity} we discuss the overall sensitivity of setups identical or similar to existing dark matter haloscopes and catalog the GW-cavity coupling coefficient for various resonant modes, GW propagation directions, and GW polarizations. Finally, in \Sec{conclusions} we conclude and give an outlook on future detection possibilities. Appendix~\ref{app:cavitymodes} contains additional details about cavity mode functions and energy densities.

\section{GW electrodynamics in the Proper Detector frame}
\label{sec:general}

In this section, we provide a detailed discussion of GW electrodynamics, paying particular attention to the role of gauge invariance. Before presenting the technical details, we give a conceptual overview of the signal strength and the process of graviton-photon conversion in the language of classical fields. As we show in the following sections, we find this formalism particularly convenient at the level of identifying optimal cavity modes and quantifying the dependence of the signal on the GW's direction of propagation. Our notation and conventions follow those of Ref.~\cite{Maggiore}. 

The GW-EM coupling is encapsulated in the Einstein-Maxwell action
\begin{equation}
\label{eq:action1}
    S = \int d^4 x ~ \sqrt{-g} \left(-\frac{1}{4} \, g^{\mu \alpha} \, g^{\nu \beta} \, F_{\mu \nu} \, F_{\alpha \beta}\right),
\end{equation}
where $F_{\mu \nu}$ is the EM field strength. To isolate the effect of a GW, we first linearize the metric as $g_{\mu\nu} = \eta_{\mu\nu} + h_{\mu\nu} +\mathcal{O}(h^2)$, where $\eta_{\mn} = \text{diag}(-1,1,1,1)$ is the flat-space metric, $h_{\mn}$ is the dimensionless GW strain, and $\order{h^n}$ denotes a quantity order-$n$ in strain $\hmn$. In the presence of a static external $B$-field $\B_0$, the action contains $\order{h}$ terms schematically of the form $\sim h \, \B \cdot \B_0$. This implies that a GW of frequency $\wg$ can generate an EM field of typical magnitude $h\, \B_0$ at the same frequency. Inside an EM cavity, this signal will ring up coherently if $\wg$ matches the cavity's resonant frequency. At the level of single quanta, this effect can be interpreted as graviton-photon mixing in a background magnetic field, known as the inverse-Gertsenshtein effect~\cite{gertsenshtein1962wave,Zeldovich:1972mn,zel1973electromagnetic}. We can also describe this effect in terms of a classical effective current, which as we show below is parametrically of size $\jeff \sim \wg \, h \, B_0$ when the cavity size is of order $L_{\rm det} \sim 1/\wg$. Because the graviton is described by a spin-2 tensor field, the direction of this effective current is non-trivially determined by the polarization of the GW. 

As mentioned above, we make use of the PD frame throughout this work. This frame utilizes so-called Fermi-normal coordinates~\cite{Fermi:1922abc,Fermi:1922def,Manasse:1963zz}, which describe GWs according to a freely-falling inertial observer (see footnote~\ref{f.freefalling}) and are written as an expansion in the proper distance from the detector's center of mass. The lowest order terms in this expansion were derived in Refs.~\cite{Manasse:1963zz,Ni:1978zz,doi:10.1063/1.524203,doi:10.1063/1.524292} and to all orders in Refs.~\cite{Marzlin:1994ia,Rakhmanov:2014noa}. As we illustrate below, the EM signals generated by GWs in resonant cavities are most simply described using such coordinates. Regardless, this computation is non-trivial when the GW wavelength $\lambda_g$ is comparable to the cavity size $L_\text{det}$, in which case the expansion parameter is $L_\text{det} / \lambda_g \sim \order{1}$ and the series expansion cannot be approximated by the first few terms~\cite{Fortini1982,Baroni:1984ptn,Flores:1986zv,Cal1987,Faroni:1992}. As far as we are aware, a closed-form expression for the metric, including terms to all orders in $L_\text{det} / \lambda_g$, has not been presented previously. In particular, we show below that resumming the metric in the PD frame is possible for a monochromatic GW of any wavelength traveling along a fixed direction.

\subsection{Analogies with Axion Dark Matter Detection}
\label{sec:axion_analogy}
Though it is not strictly necessary for the logic of the paper, it is useful at this point to make an analogy with axion-photon conversion, since this will allow us to derive a quick back-of-the-envelope estimate for the sensitivity of existing axion experiments to GWs. Indeed, the similarity of the phenomenology of axions and gravitons interacting with EM fields has been noted since the seminal paper of Raffelt and Stodolsky~\cite{Raffelt:1987im}, and the effective current formalism \cite{Wilczek:1987mv} is often used when studying axion dark matter signals in the low-frequency (quasistatic) limit~\cite{Sikivie:2013laa,Kahn:2016aff}. The Lagrangian for an axion dark matter field $a$ interacting with EM fields is $\mathcal{L} = -\frac{1}{4} \, g_{a \gamma \gamma} \, a \, F_{\mu \nu} \tilde F^{\mu \nu} = g_{a \gamma \gamma} \, a \ \E \cdot \B$, where $g_{a \gamma \gamma}$ is the dimensionful axion-photon coupling. Taking $\B = \B_0$ to be a static external $B$-field, the Lagrangian now contains the bilinear $g_{a \gamma \gamma} \, a \, \E$, which allows an axion field at frequency $\w_a$ to convert to an $E$-field that oscillates at the same frequency, with typical magnitude $g_{a \gamma \gamma} \, a \, \B_0$. This is reflected in the equations of motion for the axion and EM fields, which can be written so that the time derivative of a non-relativistic axion background field sources an effective current term $\jv_\text{eff} \supset g_{a \gamma \gamma} \, \partial_t a \, \B_0 \simeq \w_a \, \theta_a \, \B_0$  on the right-hand side of Amp\`{e}re's Law. Here, we defined the effective dimensionless field $\theta_a \equiv g_{a \gamma\gamma}a$, which will allow for a useful comparison to the GW case discussed above. Since axion dark matter is described by a a non-relativistic spin-0 field, the direction of the effective current is determined straightforwardly by the external field $\B_0$, independent of the axion.\footnote{To be more precise, the dominant coupling in the effective current for non-relativistic axions only involves the time derivative and not the gradient of the axion field; instead, for \emph{relativistic} axions~\cite{Dror:2021nyr}, the wavevector partially determines the direction of the effective current. However, since gravitons are massless, GWs are always relativistic in this sense, which is an important difference with the axion scenario.}

\begin{figure}
    \centering
    \includegraphics[width=0.7\textwidth]{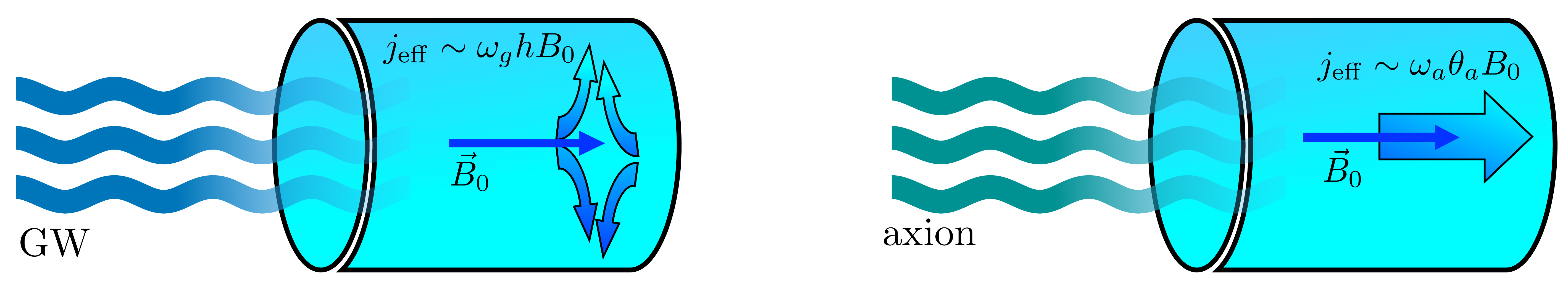}
    \caption{A cartoon illustrating the differences between GW-EM conversion (left) and axion-EM conversion (right) in the presence of an external magnetic field $\B_0$. The GW effective current is proportional to $\wg h B_0$, with a direction dependent on the GW polarization and a typical quadrupole pattern, yielding a signal field with amplitude $h B_0$. The axion effective current is proportional to $\w_a \theta_a B_0$, with a direction parallel to the external field $\B_0$, yielding a signal field with amplitude $\theta_a B_0$. The differing geometry of the effective current yields different selection rules for coupling the GW and axion to cavity modes.}
    \label{fig:setup-cartoon}
\end{figure}

A schematic illustration of this axion vs.\ GW comparison is shown in \Fig{setup-cartoon}. The effective current formalism helps elucidate the fact that the cavity modes which couple most strongly to GWs will in general be different from those excited by axions. Nonetheless, we will show below that for certain geometries, GWs do indeed have a non-zero coupling to the TM$_{010}$ cavity mode currently employed in, e.g., the ADMX and HAYSTAC axion detectors, meaning that these  experiments already have some sensitivity to GWs in their resonant frequency ranges. Momentarily ignoring very important differences in the spectral characteristics of the axion dark matter and GW fields, we can derive a conservative estimate for the sensitivity of axion dark matter experiments to coherent high-frequency GWs by comparing the respective forms of the effective currents. In particular, identifying $\theta_a \sim h$ and noting that ADMX is currently sensitive to the QCD axion parameter space, corresponding to $\theta_a \sim \text{several} \times 10^{-22}$, implies that such experiments are sensitive to similar values of the strain $h$ (as well as smaller values for GW signals that are more coherent than axion dark matter). A more precise sensitivity estimate will be provided in \Sec{sensitivity}.

Aside from the difference in cavity mode selection rules, there is a second important conceptual difference between axions and gravitons related to the role of reference frames. The axion dark matter field is assumed to have a Maxwellian speed distribution in the galactic rest frame, and moving to the laboratory frame where the cavity fields are defined is a simple Galilean boost which does not parametrically affect the signal strength. On the other hand, the large gauge freedom of linearized general relativity allows the GW signal to be computed in different reference frames, but a gauge transformation will also transform the background EM fields at the same order as the signal strength. We explore these issues in detail below.

\subsection{General Formalism}

The equations of motion can be derived straightforwardly from the action of \Eq{eq:action1}. In particular, after linearizing in $\hmn$ and integrating by parts, the $\order{h}$ terms in the action are
\be
\label{eq:action2}
S \supset %+
- \frac{1}{2} \, \int d^4 x ~ \jeff^\mu \, A_\mu
~,
\ee
where $\jeff^\mu \sim \order{h}$ is an \emph{effective} 4-current
\be
\label{eq:jeff1}
\jeff^\mu \equiv %- 
\partial_\nu \left( \frac{1}{2} \, h \, F^{\mu \nu} + h^\nu_{~ \alpha} \, F^{\alpha \mu} - h^\mu_{~ \alpha} \, F^{\alpha \nu} \right)
~,
\ee
and $h \equiv \eta^{\mn} \hmn$ is the trace of $\hmn$. It is important to emphasize that $\jeff^\mu$ is not a true 4-current; for instance, it does not transform covariantly between frames. This stands in contrast to other tensors, such as the Riemann tensor $R_{\mn \rs} \sim \order{h}$, which upon transforming between frames behaves schematically as $R_{\mn \rs} \to R_{\mn \rs} + \order{h^2}$ and is thus frame-independent to $\order{h}$ \cite{Fortini1982,Fortini:1990mr}. Hence, although $\jeff^\mu$ is $\order{h}$, it is \emph{not} invariant between frames. At a technical level, this arises from the fact that in linearized theory, the form of $\hmn$ is not unique, up to a residual gauge freedom in the choice of coordinates $x^\mu \to x^{\prime \mu} = x^\mu + \xi^\mu (x)$, where $\partial_\mu \xi_\nu \sim \order{h}$. From the usual transformation properties of a tensor we can conclude that an $\order{h^n}$ tensor transforms as $\order{h^n} \to \order{h^n} + \order{h^{n+1}}$, which implies that a covariant $\order{h}$ tensor is gauge-invariant in linearized theory. This transformation, applied to $h_{\mu\nu}$ and $F_{\mu\nu}$, can also be used to show explicitly that $\jeff^\mu$ is not invariant at $\order{h}$. That being said, $\jeff^\mu$ does couple to the EM potential as a 4-current does, and so sources signal EM fields when a GW encounters a background EM field.\footnote{In certain setups, EM signals can also arise indirectly from mechanical deformations of the experimental apparatus. Since such effects are not relevant for the experimental setups discussed in this work, we postpone this discussion to future studies.} The remainder of this section is dedicated to further illustrating this point in a manner that is largely agnostic to the particular experimental setup.

To evaluate \Eq{eq:jeff1}, we must choose a frame. The preferred frame of the laboratory is the aforementioned PD frame, which generalizes the notion of an inertial observer to curved space-time and reduces to the flat space-time metric in the $\wg \to 0$ limit~\cite{Manasse:1963zz,Misner1973,Maggiore}. Furthermore, unlike in other frames, experimental parameters, such as the applied EM fields, are naturally defined in the PD frame. In most applications of the PD frame, the metric is written to leading order in $\wg \, L_\text{det} \ll 1$, where $L_\text{det}$ is the characteristic size of the detector. However, this is not suitable for the experimental setup explored in this study, which instead requires $\wg \, L_\text{det} \sim \order{1}$. In our calculations, we thus employ the full set of higher-order terms in the metric, which has been derived in, e.g., Refs.~\cite{Marzlin:1994ia,Rakhmanov:2014noa}. Adapting these results, we find that for  $R_{\mn \rs} \propto e^{i \wg (t-z)}$ in the form of a monochromatic plane wave traveling along the $+ \hat{z}$ direction, the metric in the PD frame can be obtained exactly by resumming the full set of terms from Ref.~\cite{Marzlin:1994ia},
\begin{align}
\label{eq:hfermi}
h_{00} &= - R_{0i0j} \, x^i \, x^j \times 2 \left[ - \frac{i}{\wg z} + \frac{1 - e^{- i \wg z}}{(\wg z)^2} \right]
\nl
h_{ij} &= - \frac{1}{3} \, R_{ikjl} \, x^k \, x^l \times 6 \left[ - \frac{1 + e^{- i \wg z}}{(\wg z)^2} - 2 i \,  \frac{1 - e^{- i \wg z}}{(\wg z)^3}  \right]
\nl
h_{0i} &= - \frac{2}{3} \, R_{0jik} \, x^j \, x^k \times 3 \left[ - \frac{i}{2 \, \wg z} - \frac{e^{- i \wg z}}{(\wg z)^2} - i \,  \frac{1 - e^{- i \wg z}}{(\wg z)^3}  \right]
~,
\end{align}
where the symbol $i$ denotes a spatial index only when  appearing as a superscript or subscript of a vector or tensor; otherwise it is understood to be the imaginary unit.\footnote{\label{f.freefalling}In general, \Eq{eq:hfermi} contains additional terms that are, for instance, tied to the rotation and gravitational field of the Earth, and thus contribute slow-frequency variations to the metric that can be ignored for the experimental setups that we are interested in~\cite{Maggiore}.}  In \Eq{eq:hfermi}, the Riemann tensor $R_{\mn \rs}$ is evaluated at the spatial origin $x^ i = 0$, which is chosen to coincide with the detector's center of mass, keeping the time-dependence $R_{\mn \rs}=R_{\mn \rs}(\w_g t)$ intact. Note that the second factor in each line of \Eq{eq:hfermi} is equal to unity in the limit that~$\wg \to 0$ and so smoothly matches on to the well-known long-wavelength expressions~\cite{Maggiore}. To proceed, we take advantage of the fact that $R_{\mn \rs}$ is frame-independent to leading order in strain~\cite{Maggiore}. For convenience, we can thus choose to evaluate $R_{\mn \rs}$ in the TT frame, in which case \Eq{eq:hfermi} becomes
\begin{align}
\label{eq:hfermi2}
h_{00} &=  - \wg^2 \, h_{ab}^\TT \, x^a \, x^b  \left[ - \frac{i}{\wg z} + \frac{1 - e^{- i \wg z}}{(\wg z)^2} \right]
\nl
h_{ij} &= \wg^2 \, \Big[ \big( \delta_{i z} \, h_{ja}^\TT + \delta_{j z} \, h_{ia}^\TT \big) z \, x^a - h_{ij}^\TT \, z^2 - \delta_{iz} \, \delta_{jz} \, h_{a b}^\TT \, x^a \, x^b \Big] \, \left[ - \frac{1 + e^{- i \wg z}}{(\wg z)^2} - 2 i \,  \frac{1 - e^{- i \wg z}}{(\wg z)^3}  \right]
\nl
h_{0 i} &= - \wg^2 \, \Big( h_{ia}^\TT \, z \, x^a - \delta_{iz} \, h_{ab}^\TT \, x^a \, x^b \Big) \left[ - \frac{i}{2 \, \wg z} - \frac{e^{- i \wg z}}{(\wg z)^2} - i \,  \frac{1 - e^{- i \wg z}}{(\wg z)^3}  \right]
~,
\end{align}
where $h_{ab}^\TT \propto e^{i \wg t}$ is the GW in the TT frame evaluated at the spatial origin, and the indices $a, b = x,y$ run over the spatial components that are perpendicular to the GW's direction of propagation. Substituting \Eq{eq:hfermi2} into \Eq{eq:jeff1} yields $\jeff^\mu$ in terms of $\hmn^\TT$ for a monochromatic plane wave.\footnote{The resulting expression of $\jeff^\mu$ is quite cumbersome and not particularly illuminating with regards to the underlying physics, but for completeness it is provided in \Eq{eq:staticBz} for the case of a spatially-uniform static magnetic field.} As a result, the physical effects of the GW, as well as our projected sensitivity, will be phrased in terms of $\hmn^\TT$, which can be regarded as the strain sourced by some cosmological or astrophysical event in the TT frame, decomposed in terms of its plus- or cross-polarized components $h_+^\TT$ and $h_\times^\TT$~\cite{Maggiore}. To simplify the notation, we will drop the ``TT" superscript on the strain polarizations, $h_{+ , \times} \equiv h_{+ , \times}^\TT$, for the remainder of this work.

\subsection{Toy Example}

We conclude this section with a simple toy example calculation involving a GW impinging on a longitudinal magnetic field. As we will see below, this calculation highlights the importance of keeping track of frame-dependent aspects of the signal, some of which have been overlooked in the past literature but are necessary to preserve gauge invariance. Generally, we aim to illustrate the important point that laboratory parameters, such as the EM-field configuration or motion of a sensor, cannot be held fixed in both the PD and TT frames. At the level of this toy example, we concretely define the experimental setup, and then transform the relevant quantities into the frame of interest in order to calculate an approximately gauge-invariant observable in the form of an $\order{h}$ covariant vector (see the discussion below \Eq{eq:jeff1}).

For simplicity, we begin by considering a low-frequency plus-polarized GW of amplitude $h_+$ traveling along the $+ \hat{z}$ direction over a region of space that contains a magnetic field $\B_0$. To proceed with the calculation, we must specify a detector, two examples of which are discussed here. One such detector is an ideal conductor. In this case, we expect the oscillating EM fields generated by the GW to drive a small signal current that scales as $J_\text{sig} \propto \sigma \, h_+ \, B_0$, where $\sigma$ is the conductivity of the detector. Let us take the background magnetic field to be static, spatially uniform, and pointing along the $+\hat{z}$ direction in the TT frame, $\B_0^\text{TT} = B_0 \, \hat{z}$. Though real experimental $B$-fields will usually be static in the PD frame, this choice allows us to most easily compare to previous calculations (which work solely in TT gauge~\cite{Raffelt:1987im}) and to demonstrate the frame dependence. For concreteness, we take the position of the conducting detector to be approximately fixed in the PD frame. This is physically motivated and can be ensured by fixing the detector in the laboratory to an arbitrarily stiff mount that can resist the tidal forces induced by the GW.

It is instructive to compare the signal current $J_\text{sig}$ as derived in either the TT or PD frame. In the TT frame, the trace and divergence pieces of the metric vanish, in which case only the third term of \Eq{eq:jeff1} contributes to the effective current of the previous section, such that 
\be
j_{\text{eff}, \TT}^\mu =  %+
- \FTT^{\alpha \beta} ~ \partial_\beta h^{\TT\, \mu}_{~ \alpha}
\simeq %- 
i \wg \, \FTT^{i z} \, h^{\TT \, \mu}_{~ i}
= 0
~,
\ee
where we took the GW to be a monochromatic plane wave traveling along the $+\hat{z}$ direction, $h_{\mn}^\text{TT} \propto e^{i \wg (t - z)}$. The last equality shows that in the TT frame the effective current vanishes in the case that the background magnetic field is aligned with the GW's direction of propagation.

To illustrate the signal, we adopt from Ref.~\cite{Rakhmanov:2014noa} the coordinate transformations from the PD frame (denoted with $x^\mu$) to the TT frame (denoted as $\xTT^\mu$), which to leading order in the long wavelength limit are\footnote{In adopting the results of Ref.~\cite{Rakhmanov:2014noa}, we have switched to our convention of a GW propagating in the $+\hat{z}$-direction.}
\begin{align}
\label{eq:PDtoTT}
\tTT &\simeq t - \frac{i}{4} \, \wg \, (x^2 - y^2) \, h_+ \, e^{i \wg t}
~~,~~
\xTT \simeq x - \frac{1}{2} \, x \, (1 - i \wg z) \, h_+ \, e^{i \wg t}~.
\nl
\yTT &\simeq y + \frac{1}{2} \, y \, (1 - i \wg z) \, h_+ \, e^{i \wg t}
~~,~~
\zTT \simeq z - \frac{i}{4} \, \wg \, (x^2 - y^2) \, h_+ \, e^{i \wg t}
~.
\end{align}
\Eq{eq:PDtoTT} can be used to transform the 4-velocity $U^\mu$ of the conducting detector in the PD frame to the TT frame. Since the conductor is assumed to be stationary in the PD frame $U^\mu = (1,0,0,0)$, this implies that it moves with a small $\UTT^i \sim \order{h_+}$ velocity in the TT frame. Such movement in the stationary background magnetic field $\B_0^\text{TT}$ generates a Lorentz force, driving a small oscillating signal current. More concretely, the covariant generalization of Ohm's law $J_\text{sig}^\mu - J_\text{sig}^\nu \, U_\nu \, U^\mu = \sigma \, F^{\mn} \, U_\nu$ dictates that the signal current\footnote{Here and in the rest of the paper, we define 4-currents to be covariant vectors (such that, e.g., $D_\mu F^{\mu\nu}=J^\nu$) and not tensor densities.} is $J_\text{sig,TT}^i = \sigma \, \FTT^{i \mu} \, U_{_{\text{TT}} \, \mu} + \order{h^2}$~\cite{jackson_classical_electrodynamics_1999}, which yields
\be
\label{eq:jsigTT}
J_\text{sig,TT}^i \simeq \frac{i}{2} \, \sigma \, B_0 \, \wg \, h_+ \, e^{i \wg t} \, (y, x, 0)
~
\ee
to leading order in $\wg \times \text{length}$. Hence, as expected, we see that the effect of the GW is to drive an oscillating current in the conductor at the same frequency as the GW. The calculation proceeds in a similar manner in the PD frame, in which case the coordinates of the conductor are fixed. However, transforming the EM field-strength from the TT to PD frame shows that the EM fields pick up an oscillating component from the GW,
\be
\label{eq:EBPD}
\E \simeq \frac{i}{2} \, B_0 \, \wg \, h_+ \, e^{i \wg t} \, (y, x, 0)
~~,~~
\B \simeq \B_0^\text{TT} - \frac{i}{2} \, B_0 \, \wg \, h_+ \, e^{i \wg t} \, (x , -y , 0)
~.
\ee
From the above expression for the electric field, we see that the signal current in the PD frame, $J_\text{sig}^i = \sigma \, E^i$, agrees with the same result in the TT frame in \Eq{eq:jsigTT} (we have also checked that this agreement holds to all orders in $\wg \times \text{length}$).

Another example of a simple detector is a wire that is prepared by driving a current along the $+ \hat{z}$ direction, parallel to the applied magnetic field, such that the wire experiences zero Lorentz force in the absence of a GW. However, we expect a signal to arise in the form of an oscillating Lorentz force, distinct from the usual tidal force, when the GW sources small EM fields perpendicular to the direction of the wire. As before, we take the background magnetic field to be $\B_0^\text{TT} = B_0 \, \hat{z}$ in the TT frame, and for concreteness, we assume that the applied current density $J_0$ in the wire is constant in time in the PD frame, $J_0^\mu = (0,0,0,J_0)$. For a GW propagating in the $+\hat z$ direction, $j_{\text{eff}, \TT}^\mu = 0$, as in the previous example, and one might be tempted to conclude that no EM signal is generated. However, using the transformations of \Eq{eq:PDtoTT}, the current density in the wire picks up additional oscillating components in the TT frame, 
\be
J_{0,\text{TT}}^\mu \simeq J_0^\mu + \frac{i}{2} \, J_0 \, \wg \, h_+ \, e^{i \wg t} \, (0,x,-y,0)
~,
\ee
which induce small Lorentz forces. In particular, from the covariant form of the Lorentz force density, $f_\text{sig}^\mu = F^{\mn} J_{0 \, \nu}$, we obtain
\be
\label{eq:forceTT}
f_\text{sig,TT}^\mu \simeq - \frac{i}{2} \, J_0 \, B_0 \, \w_g \, h_+ \, e^{i \wg t} \, (0,y,x,0)
~.
\ee
In the PD frame, we calculate the force density obtained by the corrections to the EM fields in \Eq{eq:EBPD}. As expected, we find that this agrees with the TT force density of \Eq{eq:forceTT}.

From the calculations above, we see that the signal, as described by the $\order{h}$ covariant current or Lorentz force density, is invariant between the TT and PD frames. Generally, the EM signal sourced by a GW interacting with a background EM field has individual contributions that differ between frames but whose sum enters into a gauge-invariant observable. In the remaining sections, our calculation differs somewhat from the toy example presented above, in that the applied magnetic field $\B_0$ is considered to be spatially uniform and static in the PD frame (instead of the TT frame). In this case, in the TT frame the effective current vanishes for a GW propagating in the direction of $\B_0$, and instead the signal arises from the transformation of the background EM fields between the PD and TT frames. Previous studies have neglected the latter contribution by taking the background EM field in the TT frame to be the same as in the laboratory; in this case, one is led to mistakenly conclude that the signal vanishes for a magnetic field aligned with the GW's direction of propagation (see, e.g., Refs.~\cite{DeLogi:1977qe,Li:2009zzy,PhysRevD.104.023524}).

For a more general geometrical configuration, the signal in the TT frame has contributions both from the effective current and the transformation of EM fields between frames. On the other hand, for a rigid detector in the PD frame the signal solely arises from the effective current, since the background EM fields are defined with respect to the laboratory. Since the isolation of the effective current as a source simplifies the analysis for more general geometries, we work solely in the PD frame for the remainder of this work. In the next section, we develop this formalism further, by evaluating $\jeff$ and the resulting signal for an experimental setup consisting of a static magnetic field inside a cylindrical cavity.

To conclude, we note that the main lesson from our toy example is simple and well-known: we can measure only gauge-invariant quantities. Above, we computed two quantities, $J_\text{sig}^\mu$ and $f_\text{sig}^\mu$, which are frame-independent at $\order{h}$. Going beyond this, the Lorentz force, for instance, projected onto the 4-velocity of an observer $f_\text{sig}^\mu \, U_\mu = F^{\mu\nu} \, J_\nu \, U_\mu$ is frame-independent to all orders in $h$. In the remainder of the paper, we implicitly compute gauge-invariant observables, with the contraction with $U_\mu$ understood. Appropriate probes in the laboratory (and so in the PD frame), such as test-currents $J_\nu$ and observers $U_\mu$, are sensitive to the different $F^{\mu\nu}$ components that we compute below.

\section{Resonant excitation of cavities}
\label{sec:currents}

\begin{figure}
    \centering
    \includegraphics[width=0.8\textwidth]{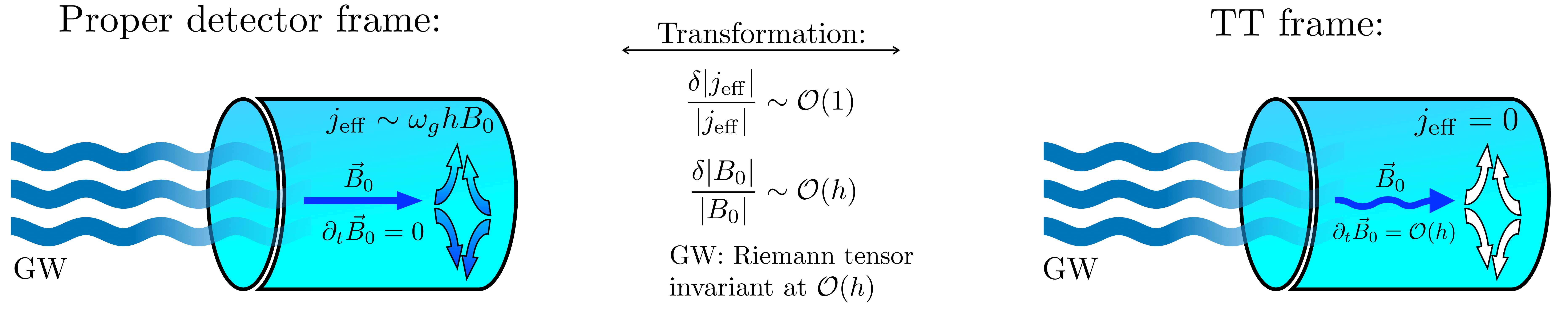}
    \caption{A heuristic sketch of our experimental setup in the PD and the TT frames, as well as the transformations between them. The EM signal arises from a GW-induced effective current in the PD frame, where the external magnetic field $\B_0$ is static. In the TT frame the effective current vanishes for specific $B$-field geometries, but there is nonetheless a signal from an $\order{h}$ oscillating correction to $\B_0$. To avoid these gauge artifacts, we work entirely in the PD frame, which corresponds to the experimental situation where the EM fields are set up in a frame where the detector and the applied $B$-field are static. }
    \label{fig:cartoon}
\end{figure}

In this section, we calculate the EM signal that arises in a resonant cavity immersed in a magnetic field $\B_0$. As emphasized above, in a realistic experimental setup, $\B_0$ is static and spatially uniform in the PD frame. Note that this is essentially the reverse of the toy example discussed above, where $\B_0$ was instead taken to be static in the TT frame. By contrast, our calculational setup, as illustrated in Fig.~\ref{fig:cartoon}, corresponds to the physical situation of turning on a static and spatially-uniform magnetic field in the PD frame.

We begin in \Sec{cavity_general} by discussing the general form of the signal  for an arbitrary cavity, which admits a decomposition of the EM fields into resonant modes~\cite{Hill}. We continue in \Sec{selection_rules} by focusing on cylindrical cavities that are already employed in searches for axion dark matter. In this case  analytic expressions for the cavity modes are tractable, and considerations of the EM signal greatly simplify if the external magnetic field is oriented along the symmetry axis of the cavity. 

\subsection{General Formalism}
\label{sec:cavity_general}

The components of the effective current $\jeff^\mu = (\rho_\text{eff} \, , \, \jveff)$ enter as additional source terms in the inhomogeneous Maxwell equations, 
\begin{eqnarray}
\nabla\cdot\E&=& \rho_{\rm eff} + \rho, \label{eq:InHomo1}\\
\nabla\times\B-\partial_t\E&=& \jveff + \jv\label{eq:InHomo2}
~,
\end{eqnarray}
where $\rho$ and $\jv$ are physical charge and current densities. The piece of $\jv$ independent of $h_{\mu \nu}$ sources, for example, the external $B$-field: $\nabla\times\B_0=\jv_0$. In the following we subtract all such zeroth-order pieces from Eq.~\eqref{eq:InHomo2} such that all fields are $\order{h}$.
On the other hand, the homogeneous Maxwell equations (Gauss' law for magnetism, $\nabla\cdot\B=0$, and Faraday's law, $\nabla\times\E+\partial_t\B=0$) are not modified by the GW. Indeed, this is because these Maxwell equations come from a topological equation of motion $dF = 0$ (where $d$ is the exterior derivative and $F$ is the field-strength two-form) which does not involve the metric $g_{\mu \nu}$. This will be of practical importance because the homogeneous Maxwell equations determine the resonant cavity modes, and thus we will see that the change in the modes induced by the tidal force of the GW affects our signal only at $\mathcal{O}(h^2)$. 

Combining \Eq{eq:InHomo2} with Faraday's law yields the standard form of the wave equation
\be
\label{eq:vectorizedHelmholtzE}
\nabla\times\nabla\times \E + \partial_t^2\E = - \partial_t\jveff - \partial_t\jv
~.
\ee
The electric field in the cavity is expanded in terms of the resonant modes $\E_n(\xv)$ as~\cite{jackson_classical_electrodynamics_1999,Hill,Collin}: 
\be
\label{eq:expansion_modes}
\E(\xv,t)=\sum_n \, e_n(t) \, \E_{n}(\xv)
~,
\ee
where $n$ indexes the various modes and $e_n$ is a dimensionless time-dependent coefficient. In general, the sum over the modes $\E_n$ includes both solenoidal ($\nabla\cdot\E_{n}=0$) and irrotational ($\nabla\times\E_{n}=0$) contributions. Since irrotational modes are not resonantly enhanced (see \App{Irrational_modes_cannot_be_enhanced}), they are omitted from our analysis below. Note also that we assume any degenerate modes have been diagonalized into orthogonal mode functions and indexed separately in the sum.

The spatial mode functions $\E_n$ satisfy the relations
\begin{eqnarray}
	\nabla^2 \, \E_{n}(\xv) &=& -\w_{n}^2 \, \E_{n}(\xv)~,
		\label{eq:lap_relation}
	\\ 
	\int_{\Vcav} \hspace{-0.3cm} d^3 \xv ~ \E_{n}(\xv)\cdot\E^*_{m}(\xv)&=& \delta_{nm}\int_{\Vcav} \hspace{-0.3cm} d^3 \xv ~ \left|\E_{n}(\xv)\right|^2
	\label{eq:orthogonality_relation_t}
	~,
\end{eqnarray}
where $\w_n$ is the resonant frequency of mode $n$ and $\Vcav$ is the volume of the cavity. The above relations are supplemented by the boundary condition $\hat{\n} \times \E_n=0$ over the surface of the cavity, where $\hat{\n}$ is the unit vector normal to the boundary. Note that formally this boundary condition is unchanged even in the presence of the GW, since it follows from the standard form of Faraday's law, as mentioned above. The boundary of the cavity can oscillate in the presence of a GW, but as we will see below this only affects the signal at $\mathcal{O}(h^2)$.

Using Eqs.~(\ref{eq:expansion_modes})--(\ref{eq:orthogonality_relation_t}), the wave equation in \Eq{eq:vectorizedHelmholtzE} can be rewritten in terms of the mode coefficients $e_n$. In the absence of external physical current sources, this gives
\begin{eqnarray}
	\Big(\partial_t^2 +\frac{\w_n}{Q_n} \, \partial_t + \w_n^2 \Big) \, e_{n}(t)=
	- \, \frac{\int_{\Vcav} \hspace{-0.2cm} d^3 \xv ~ \E_n^* \cdot \partial_t \jveff}{\int_{\Vcav} \hspace{-0.2cm} d^3 \xv ~ |\E_n|^2}
	~,
	\label{eq:eqofmotion_s_t}
\end{eqnarray}
where the mode-dependent quality factor $Q_n$ appears from losses in the cavity. In principle, $\w_n$ and $\Vcav$ appearing in the above equation are perturbed by the tidal force of the GW at $\order{h}$. However, if in the absence of the GW there is no power in the cavity modes, this effect corrects our signal only by terms of $\order{h\, e_n}$ and $\order{h \, \jeff}$, both of which are $\order{h^2}$.

Taking the GW to be monochromatic and on resonance with the cavity mode, i.e., $\jveff(\xv,t)= e^{i\wg t} \, \jveff(\xv)$ where the GW frequency is $\wg \simeq \w_n$, the solution to \Eq{eq:eqofmotion_s_t} is enhanced by the large quality factor $Q_n \gg 1$. The steady-state form of the excited signal electric field $\E_\text{sig} = e_n \, \E_n$ is then given by
\be
\label{eq:Esigj}
\E_\text{sig}(\xv, t) =
	- \, \frac{\int_{\Vcav} \hspace{-0.2cm} d^3 \xv^\prime ~ \E_n^* \cdot \jveff}{\int_{\Vcav} \hspace{-0.2cm} d^3 \xv^\prime ~ |\E_n|^2} ~ \frac{Q}{\wg} ~ \E_n(\xv) \, e^{i \wg t}
	~,
\ee
where for convenience, we will drop the subscript from the quality factor, $Q \equiv Q_n$, from now on. Note that if there are multiple degenerate modes at $\wg$, we must sum over all such modes with their own time functions $e_n(t)$ to obtain the total signal field; we will see such an example for cylindrical cavities in \Sec{selection_rules} below. Since the effective current scales as $\jeff \propto \wg^2$ in both the high- and low-frequency limits (see Sec.~\ref{sec:selection_rules}), we decompose it into a dimensionful amplitude and two spatially-dependent dimensionless functions $\jplus (\xv)$ and $\jcross (\xv)$ that describe the spatial profile and polarization of the GW signal, as follows:
\be
\label{eq:jeffscale}
\jveff (\xv) \equiv B_0 \, \wg^2 \, \Vcav^{1/3} \, ( h_+ \, \jplus(\xv) +  h_\times \, \jcross(\xv) )
~.
\ee
The steady-state signal power delivered to the cavity can then be expressed as
\be
\label{eq:Psiggen}
P_\text{sig} = \frac{\wg}{2Q} \, \int_{\Vcav} \hspace{-0.2cm} d^3 \xv ~  |\E_\text{sig}|^2
~,
\ee
which follows from the energy density stored in resonant cavity modes (see \App{energy_density}). Note from \Eq{eq:Esigj} that when both polarizations in $\jveff$ are present, the power will generically have cross-terms between the two polarizations. However, for simplicity of presentation, we will often consider the case where only one of the polarizations is present at a time. In this case, we can can define a dimensionless coupling coefficient
\be
\label{eq:formfactor}
\eta_n \equiv \frac{\Big| \int_{\Vcav} \hspace{-0.2cm} d^3 \xv ~ \E_n^* \cdot \jpc \Big|}{\Vcav^{1/2} \, \big( \int_{\Vcav}  \hspace{-0.2cm} d^3 \xv ~ |\E_n|^2 \big)^{1/2}}
~,
\ee
in terms of which the signal power can be written as
\be
\label{eq:Psig}
P_\text{sig} = \frac{1}{2} \, Q \, \wg^3 \, \Vcav^{5/3} \, (\eta_n \, \hnorm \, B_0)^2
~,
\ee
where $h_0 = h_+$ or $h_\times$ is the magnitude of the GW strain. Eq.~\eqref{eq:Psig} illustrates the scaling of the power in a transparent way; when both polarizations are present, the power should be computed directly from Eqs.~(\ref{eq:jeffscale}) and~(\ref{eq:Psiggen}). In the next section, we evaluate \Eq{eq:formfactor} for different configurations, corresponding to various cavity modes, GW polarizations, and orientations of the GW's direction of propagation with respect to the symmetry axis of the cavity.

\subsection{Selection Rules for Cylindrical Cavities}
\label{sec:selection_rules}

The discussion in the previous section is valid for cavities of any shape. In the following, we focus specifically on cylindrical cavities in part because existing experiments use this geometry. For concreteness, we will consider such a cavity of equal radius and length, $R_{\rm det} = L_{\rm det}$. The solenoidal modes of a cylindrical cavity are classified into transverse magnetic (TM) and transverse electric (TE) modes. For cylindrical cavities, the generic mode number $n$, introduced in the previous section, is represented by three integers $n=(m,n,p)$ and a $\pm$ index, where $m,~n,$ and $p$ stand for the azimuthal, radial, and longitudinal mode indices, and the $\pm$ indexes a pair of degenerate modes for $m \neq 0$ with distinct azimuthal dependence.
The explicit form of the mode functions, along with expressions for the corresponding resonance frequencies $\w_{mnp}$, are provided in \App{solenoidal_modes}.

\begin{figure}[t!]
	\centering
	\includegraphics[width=0.49\textwidth]{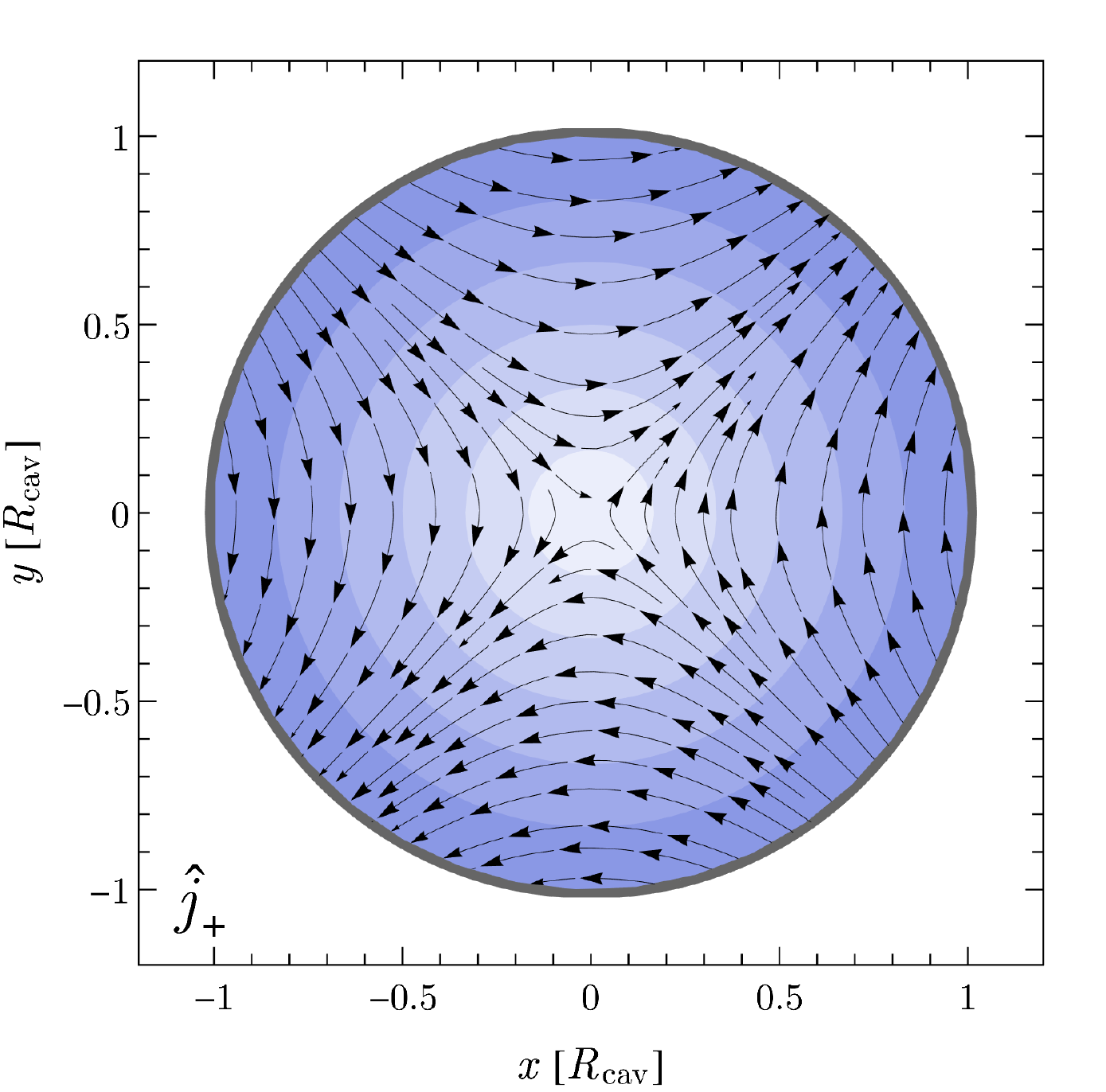}
	\includegraphics[width=0.49\textwidth]{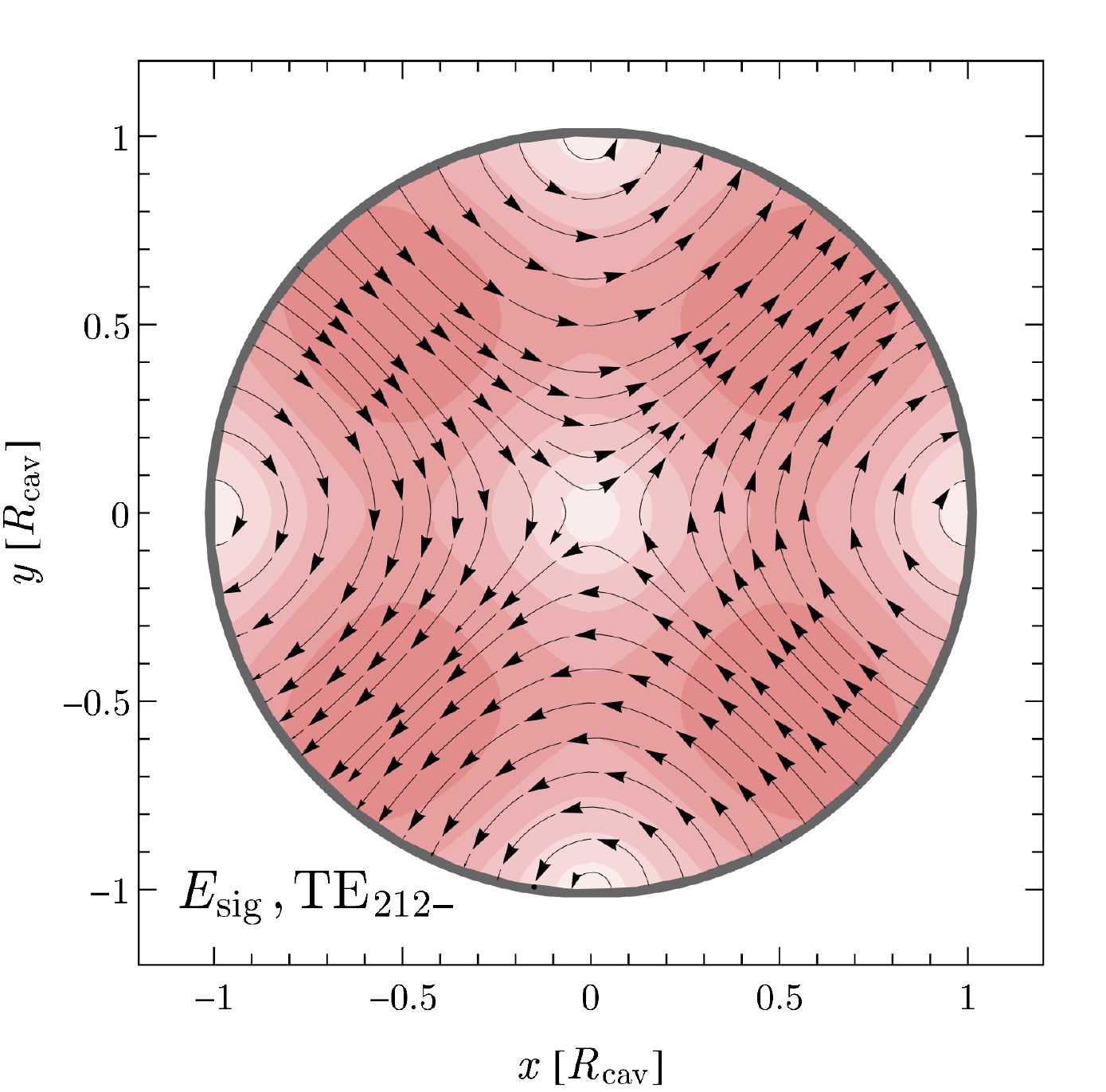}
	\caption{A cross-sectional view of the dimensionless effective current $\jplus$ (left) and electric field of the $\text{TE}_{212-}$ signal-mode $\E_\text{sig}$ (right). The black arrows denote the direction of $\jplus$ or $\E_\text{sig}$ projected onto the $xy$-plane (in units of the cavity radius $R_\text{cav}$) near $z = 0$ (corresponding to the cavity's center of mass) where the symmetry axis of the cylindrical cavity, the applied magnetic field, and the gravitational wave's direction of propagation are chosen to align with the $z$-axis (not shown). The lighter/darker colored regions correspond to smaller/larger magnitudes of $\jplus$ or $\E_\text{sig}$. The geometrical coupling between the gravitational wave and the cavity (encapsulated by the coupling coefficient $\eta_n$) is $\order{1}$, as can be inferred from the similar spatial profiles of $\jplus$ and $\E_\text{sig}$.}
	\label{fig:jeff}
\end{figure}

As illustrated by the coupling coefficient of \Eq{eq:formfactor}, it is advantageous to have a large overlap between the electric field mode  and the GW-induced effective current $\jveff$. A cross-sectional view of $\jplus$ is shown in the $xy$-plane in the left-panel of \Fig{jeff}, taking both the GW's direction of propagation and the applied magnetic field $\B_0$ to be along the cavity axis. We define our coordinate system such that the center of the cavity is located at the origin and the cavity axis is along the $z$-axis. The black arrows illustrate the direction of $\jplus$, while the lighter/darker colored regions correspond to smaller/larger magnitudes $|\jplus|$. The effective current exhibits a quadrupolar structure in the $xy$-plane, due to the spin-2 nature of the gravitational field (see the discussion below). An optimally-coupled cavity mode is one that possesses the same spin-2 structure along the azimuthal direction. This is shown in the right-panel of \Fig{jeff}, which displays the vector field for the electric component of the $\text{TE}_{212-}$ mode. As is evident by comparing both panels of \Fig{jeff}, the effective current and the electric field of the $\text{TE}_{212-}$ mode have a similar spatial dependence and therefore a large overlap, as quantified by \Eq{eq:formfactor}. In \Sec{evaluation_coupling_coefficient}, we confirm this statement by numerically evaluating the coupling coefficient $\eta_n$ for different cavity modes and find that $\eta_n \sim \order{0.1}$ is generic for various geometrical configurations, including this one. 

For this specific example, the direction of the GW and the applied magnetic field preserve the cylindrical symmetry of the cavity. As a result, the form of $\jeff^\mu$ vastly simplifies, and one can derive simple cavity mode selection rules. This is made manifest by rewriting $\jeff^\mu$ in cylindrical coordinates, i.e., $\jeff^\mu = (\rho_\text{eff}\, ,\, \jeff^r \, , \, \jeff^\phi \, , \, \jeff^z)$, such that
\begin{align}
\label{eq:staticBz}
\jeff^\mu 
= %+
- \frac{B_0  \, \wg^2 \, r}{6\sqrt{2}} \, e^{i \wg t} \,
\Big( 0  ~,~ i e^{- 2i\phi} \, h_{+2}  - i e^{2i\phi} \, h_{-2} ~,~ e^{- 2i\phi} \, h_{+2} + e^{2i\phi} \, h_{-2} ~,~ 0 \Big)
\times f(\wg z)
~,
\end{align}
where
\be
f(x) \equiv -3 - 6 i \, x^{-1} - 12 \, e^{-i x} \, x^{-2} - 12 i \, (1 - e^{-i x}) \, x^{-3}
\ee
is a dimensionless function that obeys $\lim_{x \to 0}f(x) = 1$ and $\lim_{x \to \infty}f(x) = -3$.  Above, we have introduced the GW helicity components $h_{\pm 2} \equiv (h_+ \pm i h_\times) / \sqrt{2}$, which transform under a rotation by $\Delta \phi$ about the cavity/GW axis as $h_{\pm 2} \to e^{\pm 2 i \Delta \phi} \, h_{\pm 2}$. From the explicit form of the current in \Eq{eq:staticBz}, we can understand our choice of normalization for $\jveff$ in \Eq{eq:jeffscale}. Indeed, the spatial components of the current scale as $\jveff \propto \hnorm \, B_0 \, \w_g^2 \, r$, where the typical size of $r$ is $\Vcav^{1/3}$. Since resonant modes parametrically have $r \sim 1/\wg$, this recovers the naive scaling $\jeff \sim h_0 B_0 \wg$ presented in \Sec{general}.

For a fixed GW polarization, the effective current in \Eq{eq:staticBz} is only invariant under rotations about $z$ for $\phi \to \phi + q\pi,~q\in \mathbb{Z}$, reflecting the spin-2 nature of the tensor field $h_{\mu \nu}$.\footnote{Recall that in general, spin-$j$ fields are minimally invariant under rotations $\phi \to \phi + 2\pi/j$, the famous example being the ``Dirac plate trick'' which provides a human-scale example of a system only invariant under rotations by $4\pi$~\cite{newman1942string}.} The background field $\B_0$, being uniform and aligned with the cylindrical axis, is a scalar under the cylinder's $U(1)$ rotational symmetry. The EM fields of a cylindrical cavity can be decomposed into a sum over modes with definite transformations under the cylindrical $U(1)$ symmetry, $\E \propto e^{\pm i m \phi}$ (see \App{solenoidal_modes}). Thus, from \Eq{eq:formfactor}, only modes with $m=\pm2$ couple to the effective current for this particular example. This matches the intuition garnered from \Fig{jeff}, where the angular overlap of $\jveff$ with an $m = 2$ mode is visually apparent. It is also instructive to apply this argument to the case where the GW remains coaxial with the cylinder, while $\B_0$ is not aligned with the cavity axis. In this example, the direction of the applied magnetic field does not preserve the symmetry of the cavity. Since $\B_0$ is a pseudo-vector under the cavity's $U(1)$, along with the spin-2 GW, this yields an effective current with a spin-1 component, which can couple to $m = \pm 1$ cavity modes. The coupling coefficient for such geometries is evaluated numerically in \Sec{evaluation_coupling_coefficient}.

\section{Sources}
\label{sec:sources}

Before moving on to calculating the sensitivity of an EM cavity to a generic GW in \Sec{sensitivity}, we first briefly survey the kinds of sources which could give rise to spatially-localized (on the sky) high-frequency monochromatic GWs. Our main examples are mergers of sub-solar mass objects -- including primordial black holes (PBHs) and other exotic compact objects -- and boson clouds from PBH superradiance. As we will see below, for binary mergers to generate inspiral signals in the GHz regime, the merging objects must be much lighter than a solar mass. Exotic compact objects~\cite{Giudice:2016zpa} can be considerably lighter than a solar mass and therefore can emit GWs at high frequencies. Examples of such exotic compact objects are boson and fermion stars~\cite{Palenzuela:2007dm,Giudice:2016zpa,Palenzuela:2017kcg,Helfer:2018vtq}, gravitino stars~\cite{Narain:2006kx}, gravistars~\cite{Mazur:2004fk}, and dark matter blobs \cite{Diamond:2021dth}. In general, the GW waveforms generated by merging exotic compact objects and PBHs are distinct, though this difference is small when the orbital radius is much larger than the spatial size of the merging objects; for simplicity, in our analysis we treat the merging objects as point-like. 

The frequency of the GW emitted by a merging binary increases during the inspiral phase as the two compact objects approach each other. There is an upper bound on the frequency of GWs emitted by a binary in a quasi-circular orbit in the weak-field limit, corresponding to the GW frequency at the innermost stable circular orbit (ISCO). For a binary consisting of compact objects of equal mass $M_b$, at sufficiently early times such that the orbital radius $r_b$ is greater than the ISCO radius $r_\text{ISCO} = 12 G_N M_b$, the GW frequency evolves as $\wg \simeq 14 \ \text{GHz} \times (10^{-6} \, M_\odot / M_b) \, (r_\text{ISCO} / r_b)^{3/2}$~\cite{Maggiore}. Thus, only very light binaries, such as sub-Earth mass PBHs~\cite{Hawking:1971ei,Khlopov:2008qy,Carr:2021bzv}, can generate GW signals in the GHz regime well before reaching the ISCO.

However, such GW signals are highly transient near the ISCO, as the frequency of the emitted GW evolves in time according to $d \wg / dt \propto (M_b / r_b)^{11/6}$, which can limit the sensitivity of resonant experiments. In particular, GWs from the merger of light binaries can typically only resonantly excite a cavity for a short amount of time that decreases for heavier masses. This is quantified by the number of orbital cycles $\mathcal{N}_\text{cyc}$ a binary spends emitting GWs within the resonator bandwidth $\wg / Q$, given by $\mathcal{N}_\text{cyc} \sim 10^{-3} \times (10^{-6} \, M_\odot / M_b)^{5/3} \, (10^5/Q)$ for $\wg \sim \text{GHz}$. Requiring that a typical cavity is fully rung up by the GW signal, i.e., $\mathcal{N}_\text{cyc} \gtrsim Q \sim 10^5$, requires that $M_b \lesssim 10^{-11} \ M_\odot$, which, amusingly, is a scenario where PBHs could constitute an $\order{1}$ fraction of the cosmological dark matter abundance~\cite{Niikura:2017zjd, Montero-Camacho:2019jte,Smyth:2019whb,Carr:2020gox}. The expected strain from inspirals at a distance $D$ from Earth is
\be
h_0 \sim 10^{-29} \times \bigg(\frac{1\ \text{pc}}{D}\bigg)\bigg( \frac{M_b}{10^{-11}\, M_\odot}\bigg)^{5/3} \bigg( \frac{\wg}{1 \ \text{GHz}} \bigg)^{2/3} \ .
\label{eq:strain_PBH_inspiral}
\ee
The best case strain can be estimated under the assumption that the PBHs are 100\% of the dark matter and all PBHs are paired in binaries. In this case we get $D=10^{-3}$ pc and hence a best case sensitivity of  $h_0 \sim 10^{-26}$. A more realistic study~\cite{Franciolini:2022htd} finds that the average distance where one expects one merger event per year is $D=0.21~{\rm kpc}~\left(\frac{M_B}{10^{-11}\, M_\odot}\right)^{\frac{1}{3}}$. Note hat the merger rate is used as an approximation for the number of events where the GHz frequency band is crossed by a source nearby. Plugging the more realistic value for the distance $D$ into eq.~\eqref{eq:strain_PBH_inspiral} yields $h_0=10^{-31}\,\left(\frac{M_b}{10^{-11}\,M_\odot}\right)^{\frac{4}{3}}\,\left(\frac{\wg}{1\,{\rm GHz}}\right)^{\frac{2}{3}}$.

The experimental setup we discuss in Sec.~\ref{sec:sensitivity} below has parametrically reduced sensitivity to sources that only spend a much smaller number of cycles exciting the cavity, which makes detecting heavier binaries (with correspondingly larger strain) difficult. On the other hand, since transient mergers sweep through a large range of frequencies, the resultant GWs will hit multiple resonant frequencies of the cavity. Therefore, unlike searches for axion dark matter, one does not have to scan different cavity frequencies by tuning the cavity when looking for GWs from such signals; the frequency sweeping is done by the merger itself.

High-frequency GWs can also arise from the annihilation of boson clouds generated by black hole superradiance~\cite{Ternov:1978gq,Zouros:1979iw,Arvanitaki:2009fg,Arvanitaki:2010sy,Detweiler:1980uk,Yoshino:2013ofa,Arvanitaki:2014wva,Brito:2014wla,Brito:2015oca,Zhu:2020tht}. For instance, bosons of mass $m_a \sim \mu \text{eV} \times (10^{-4} \ M_\odot / M_\text{PBH})$ accumulate in large numbers outside a PBH of mass $M_\text{PBH}$. When such bosons annihilate into gravitons in the background gravitational field, the frequency of the emitted GW is $\wg = 2 m_a \sim \text{GHz} \times (m_a / \mu \text{eV})$. Thus, if such PBHs and sub-eV bosons \emph{both} exist, superradiant clouds emit very high-frequency GWs. From the point of view of the experimental signals discussed here, the advantage of this GW source is that the associated waveform is monochromatic and coherent over very long timescales. The coherence time is limited by the change in gravitational potential energy of the boson cloud as it annihilates, which leads to a small positive drift in frequency, $d \wg / dt >0$~\cite{Zhu:2020tht}. This drift is small, such that the signal is effectively coherent for much longer than the cavity ring-up time $\sim Q/\wg$. However, the masses of such hypothetical bosons and PBHs are unknown, so unlike the case of a transient inspiral signal, a scanning strategy must be implemented. The expected strain of such signals arising from a PBH a distance $D$ from Earth is~\cite{Arvanitaki:2012cn}

\be
h_0\sim 10^{-27} \times \bigg(\frac{10 \ {\rm kpc}}{D}\bigg)\bigg(\frac{M_{\rm PBH}}{10^{-4} \, M_\odot}\bigg)
~,
\ee
where we have used that the fraction of PBH mass that the axion cloud carries is $\epsilon=10^{-3}$ and $\frac{\alpha}{\ell}=\frac{GM_bm_a}{\ell}=0.5$, where $G$ is the gravitational constant and $\ell$ is the orbital quantum number.

In agreement with all current constraints on PBH dark matter we can assume that PBHs constitute $1$\% of the dark matter density. 
This enables us to estimate the average distance to be $D\approx 1\,$pc from what we estimate $h_0\approx 10^{-23}$. Note that this is a best case scenario and the actual strain might be orders of magnitude worse~\cite{Franciolini:2022htd} because we have not taken into account how the PBHs attained the spin which is necessary to generate the boson clouds via the superradiance mechanism.

While we do not focus heavily on non-coherent sources in this paper, there are also interesting sources of high-frequency stochastic GWs. The most prominent examples include GWs arising from first-order phase transitions~\cite{Witten:1984rs,Hogan:1986qda}, cosmic strings~\cite{Damour:2000wa,Damour:2001bk}, inflation~\cite{Grishchuk:1975abc,Starobinskii:1979abc,RUBAKOV1982189,1983PhLB..125..445F}, preheating~\cite{Khlebnikov:1997di,Lozanov}, and the thermal plasma (the so-called Cosmic Gravitational Microwave Background)~\cite{Ghiglieri:2015nfa,Ghiglieri:2020mhm,Ringwald:2020ist}; for a review see, e.g., Ref.~\cite{Caprini:2018mtu}. We briefly discuss the detection of stochastic GWs with EM cavities below and will return to this issue in future work.

\section{Sensitivity estimates and cavity coupling coefficients}
\label{sec:sensitivity}

In this section, we apply the formalism developed in \Sec{currents} to a concrete experimental setup. We begin in \Sec{SNR} with a general discussion of the sensitivity of a resonant cavity to coherent or stochastic GW sources, assuming an optimal cavity-GW coupling coefficient $\eta_n \sim \order{0.1}$, and illustrate how this sensitivity scales with the assumed experimental parameters. In \Sec{evaluation_coupling_coefficient}, we then examine the angular dependence of $\eta_n$ with respect to the direction of an incoming monochromatic GW for different cavity modes and GW polarizations. We find that for various cavity modes, $\eta_n$ is $\order{0.1}$ over a large fraction of solid angle, including the $\text{TM}_{010}$ and $\text{TM}_{020}$ modes, which are already employed by existing axion dark matter experiments, such as ADMX and ORGAN. The strong dependence of the coupling coefficient on the incident direction of the GW makes a plausible case for directional detection.

We note that our estimates are purely based on the signal induced by the direct coupling of the GW to photons that we described in the previous sections. In principle, the GW can also induce other effects. We have already discussed how deformations of the cavity walls are not observable in these setups, since they generate an EM signal at $O(h^2)$. The GW can in principle also induce a relative motion between the laboratory apparatus generating the background $B$-field and the cavity itself. This relative motion can produce an additional oscillating EM field in the rest frame of the cavity. The resulting $O(h)$ oscillating component of
the background electromagnetic field is shielded by the cavity itself (even in the absence of additional electromagnetic shielding). In practice only the static component penetrates the cavity walls efficiently and can produce an $O(h)$ signal within the cavity after interacting with the wave. Thus, any electromagnetic “signal” generated by deformations of the external B-field source generate a signal only at $O(\epsilon \times h) \ll h$ where $\epsilon$ depends on the electromagnetic shielding of the cavity. 
An alternative justification for ignoring such effects is that the time-dependent electromagnetic fields generated by the GW interacting with the external electromagnetic source can only couple to the electromagnetic modes of the cavity by driving currents along the cavity walls. Such currents do not resonantly couple to the electromagnetic modes of the cavity
since, by definition, the electric components of these modes vanish at the cavity wall.

\subsection{Sensitivity Estimate}
\label{sec:SNR}

The signal power $P_\text{sig}$ due to a coherent GW on resonance with the cavity is given by \Eq{eq:Psig}. The signal-to-noise ratio (SNR) is then given by the Dicke radiometer equation as
\begin{equation}
\label{eq:SNR1}
    {\rm SNR} \simeq \frac{P_\text{sig}}{T_{\rm sys}} \,  \sqrt{\frac{t_\text{int}}{\Delta \nu}}
    ~,
 \end{equation}
where $T_{\rm sys}$ is the effective noise temperature, $t_\text{int}$ is the measurement integration time, and $\Delta \nu$ is the signal frequency bandwidth.  The sensitivity is estimated by taking $\text{SNR} \gtrsim 1$, which after using \Eqs{eq:Psig}{eq:SNR1} yields
\be
\label{eq:sensitivity_estimate}
\hnorm \gtrsim 3 \times 10^{-22} \times \bigg( \frac{1 \ \text{GHz}}{\w_g / 2 \pi} \bigg)^{3/2} \bigg( \frac{0.1}{\eta_n} \bigg) \bigg( \frac{8 \ \text{T}}{B_0} \bigg) \bigg( \frac{0.1 \ \text{m}^3}{\Vcav} \bigg)^{5/6} \bigg( \frac{10^5}{Q} \bigg)^{1/2} \bigg( \frac{T_\text{sys}}{1 \ \text{K}} \bigg)^{1/2} \bigg( \frac{\Delta \nu}{10 \ \text{kHz}} \bigg)^{1/4} \bigg( \frac{1 \ \text{min}}{t_\text{int}} \bigg)^{1/4}
~,
\ee
where we have adopted experimental parameters similar to those of ADMX~\cite{ADMX:2019uok}. Recent advances in superconducting cavity technology suggest that achieving $Q = 10^{7}$ with $B_0 = 6$~T may be possible in the near future~\cite{Posen-talk}, and of course a longer integration time is possible for a dedicated GW search.

\begin{figure}
	\centering
	\includegraphics[width=0.8\textwidth]{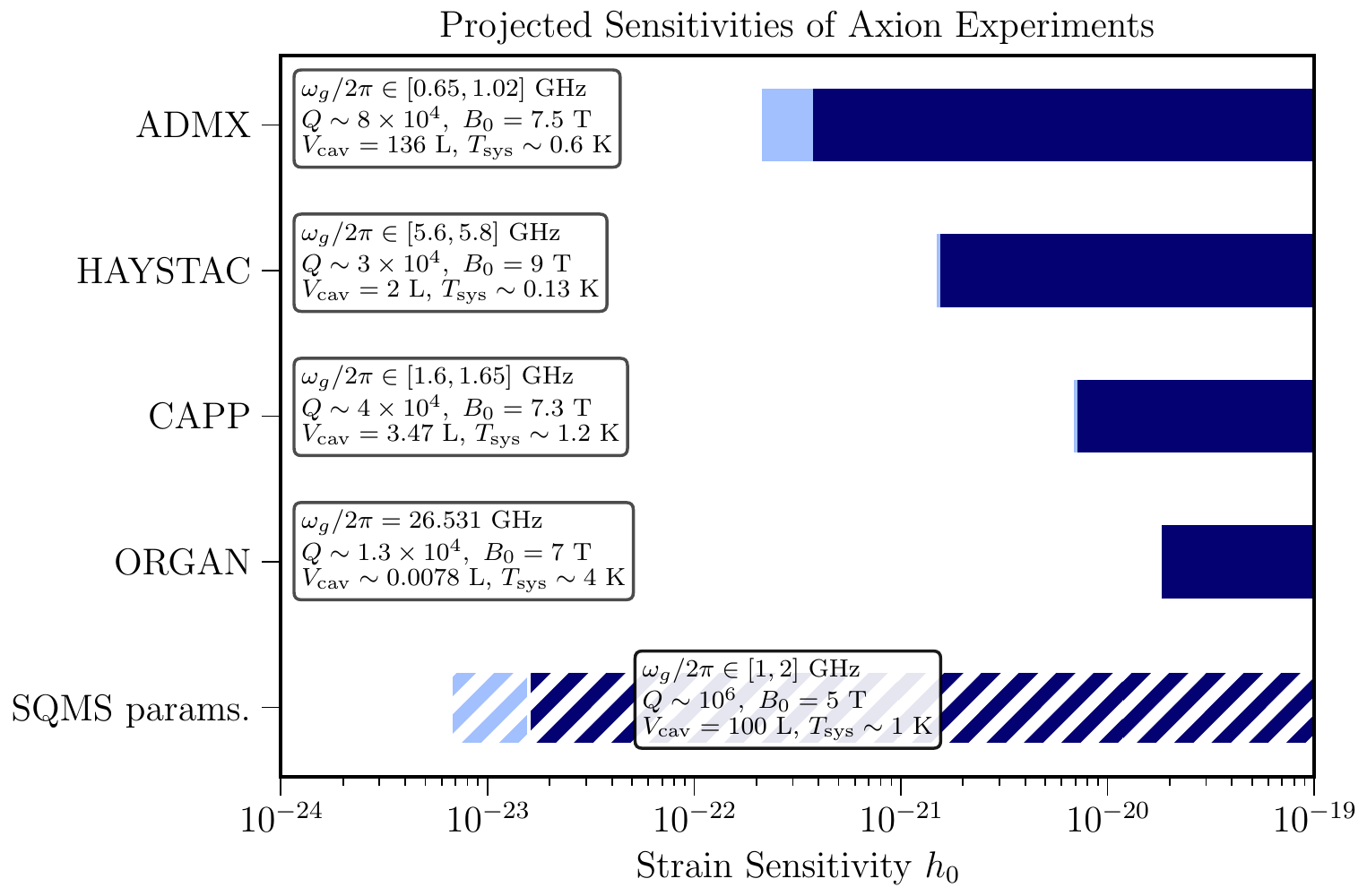}
	\caption{Projected sensitivity of axion experiments to high-frequency GWs, assuming an integration time of $t_{\rm int} = 2 \text{ min}$ for ADMX, HAYSTAC and CAPP, $t_{\rm int} = 4 \text{ day}$ for ORGAN, and $t_{\rm int} = 1 \text{ day}$ for the SQMS parameters. These integration times are characteristic of data-taking runs in each experiment. The GW-cavity coupling coefficient is fixed to $\eta_n = 0.1$ for each experiment, and the signal bandwidth $\Delta \nu$ is conservatively fixed to the linewidth of the cavity. Dark (light) blue regions indicate the sensitivity at the lowest (highest) resonant frequency of the tunable signal mode. For ADMX~\cite{ADMX:2018ogs,ADMX:2019uok,ADMX:2021nhd}, HAYSTAC~\cite{HAYSTAC:2018rwy}, and CAPP~\cite{Lee:2020cfj}, the signal mode is TM$_{010}$, but for ORGAN~\cite{McAllister:2017lkb} the signal mode is TM$_{020}$. The system temperature $T_{\rm sys}$ defining the thermal noise floor of each experiment is given in the figure, along with relevant experimental parameters including the loaded cavity quality factor $Q$. 
	}
	\label{fig:sensitivity}
\end{figure}

In a realistic setup, the signal bandwidth $\Delta \nu$ will be determined by a combination of factors. For instance, $\Delta \nu$ is bounded from below by, e.g., the intrinsic frequency spread of the GW source or the drift of the cavity resonant frequency, and it is bounded from above by the cavity bandwidth $\sim \wg/(2 \pi Q)$. Our conservative benchmark of $\Delta \nu = 10 \ \text{kHz}$ corresponds to the cavity bandwidth for $Q = 10^5$ and $\wg/2\pi = 1 \ {\rm GHz}$, similar to that of the ADMX cavity. The fundamental lower bound on the bandwidth is given by the frequency resolution $\Delta \nu \gtrsim 1/t_\text{int}$, which is saturated for a sufficiently monochromatic source with infinite coherence time in a phase-sensitive measurement scheme such as a lock-in amplifier. 
In this case, the SNR scales linearly with $t_\text{int}$, resulting in a sensitivity of
\be
\label{eq:sensitivity_estimate_coherent}
\hnorm \gtrsim 1 \times 10^{-23} \times \bigg( \frac{1 \ \text{GHz}}{\w_g / 2 \pi} \bigg)^{3/2} \bigg( \frac{0.1}{\eta_n} \bigg) \bigg( \frac{8 \ \text{T}}{B_0} \bigg) \bigg( \frac{0.1 \ \text{m}^3}{\Vcav} \bigg)^{5/6} \bigg( \frac{10^5}{Q} \bigg)^{1/2} \bigg( \frac{T_\text{sys}}{1 \ \text{K}} \bigg)^{1/2}  \bigg( \frac{1 \ \text{min}}{t_\text{int}} \bigg)^{1/2}
~.
\ee
Indeed, this technique is already used in the GW detection community in the form of ``matched filtering''~\cite{Maggiore}, which ensures that the SNR scales as $t_\text{int}$ as long as the precise waveform of the signal is known.

The sensitivity estimate from \Eq{eq:sensitivity_estimate} is illustrated in \Fig{sensitivity} for various existing axion experiments, as well as for superconducting cavities being developed at the SQMS Center at Fermilab which can support both large $B$-field and high $Q$ factors~\cite{Posen-talk}. Note that the sensitivity is generally weaker at higher frequencies because the cavity volume typically scales with the signal mode frequency as $\Vcav \propto \wg^{-3}$. After taking the volume scaling into account, we find that the sensitivity to $\hnorm$ is proportional to $\wg$, which degrades linearly at large frequencies. One might overcome this by using higher modes if the coupling coefficient for such modes does not decrease (discussed further below), though at very high mode numbers the $Q$ factor will tend to decrease and mode-crossings make isolating the signal mode difficult. Alternatively, one could consider a multiplexing strategy with $N$ cavities each of a fixed volume $\Vcav$, in which case the sensitivity would improve as $1/\sqrt{N}$. Indeed, this sort of strategy is being pursued by ADMX for their small ``sidecar'' cavities~\cite{ADMX:2018ogs}. A similar approach could be implemented in multi-cell cavities of the type typically used for RF acceleration~\cite{posen2021advances}.

Although not the main focus of this study, we conclude this subsection with a brief discussion of stochastic GW signals and leave a more detailed investigation to  future work. A stochastic GW background is described by the average strain power per unit frequency, i.e., the strain power spectral density $S_h(\w)$.  For a stochastic GW background of cosmological origin, this determines the energy density in GWs per logarithmic frequency, normalized by the critical energy density today, as $\Omega_g (\wg) \sim \wg^3 \, S_h (\wg) / H_0^2$, where $H_0$ is the present-day Hubble expansion parameter. Such a non-coherent signal appears as an additional noise source in the detector, such that the SNR is given by the ratio of the signal and noise power spectral densities, independent of the integration time~\cite{Maggiore,Grishchuk:2003un}. For the experimental setup described here, this amounts to
\be
\label{eq:SNRstochastic}
\text{SNR} \sim Q \, \wg  ~ \eta_\text{stoch}^2 \, B_0^2 \, \Vcav \, S_h (\wg) / T_\text{sys}
~,
\ee
where $\eta_\text{stoch}^2$ is the effective stochastic GW-cavity coupling coefficient. Although a stochastic signal does not resonantly excite the cavity, the SNR in \Eq{eq:SNRstochastic} scales linearly with the quality factor since larger values of $Q$ correspond to suppressed intrinsic thermal fluctuations. Experimental parameters similar to those adopted above lead to projected sensitivities that are cosmologically meaningless $\Omega_g \gg 1$, although our setup may be applicable to stochastic sources of non-cosmological origin, such as populations of merging PBHs.

\subsection{Evaluation of the Coupling Coefficient}\label{sec:evaluation_coupling_coefficient}

\begin{figure}
	\centering
	\includegraphics[width=0.49\textwidth]{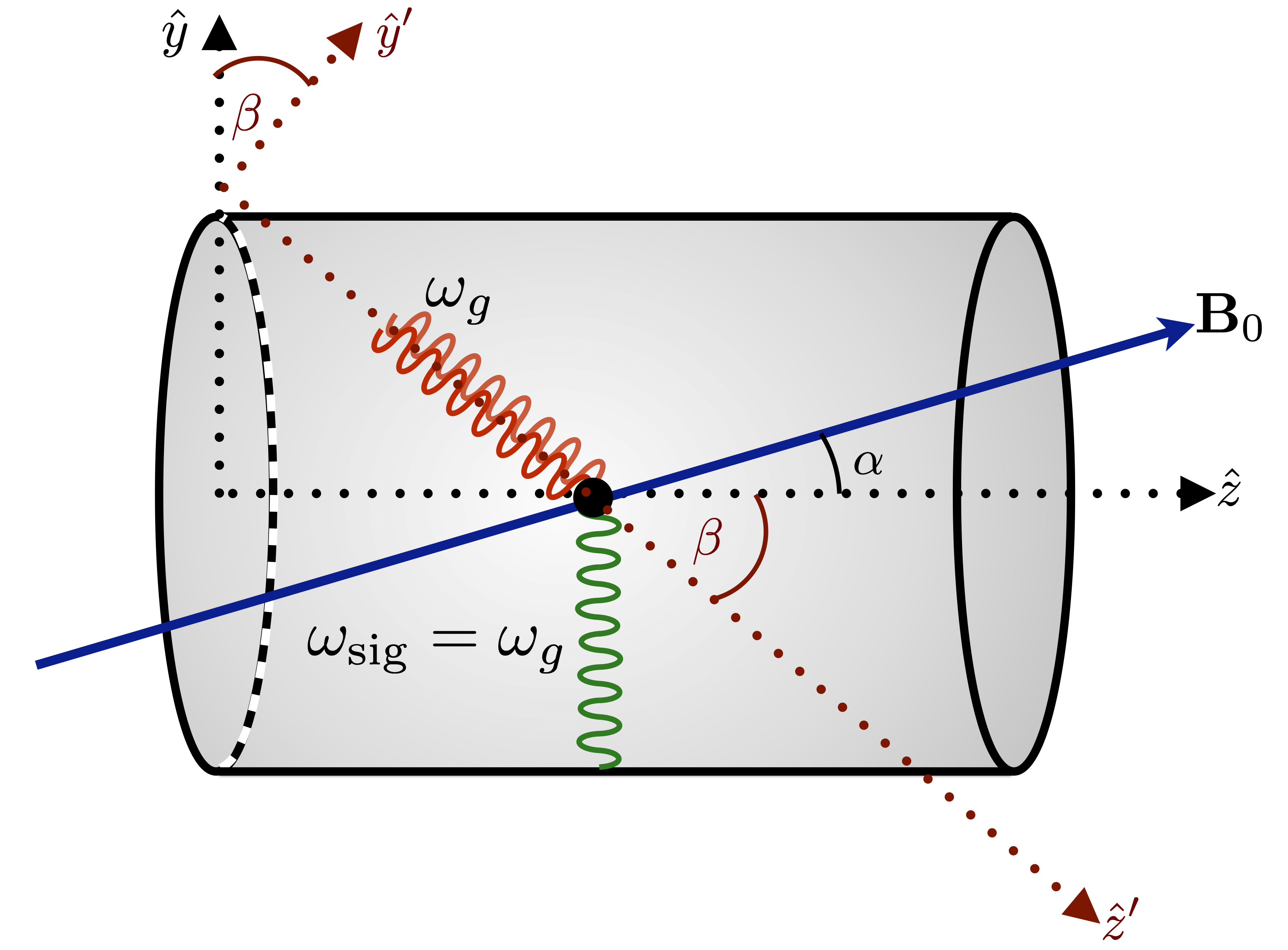}
	\caption{Geometry of a GW incident on a cylindrical cavity from an arbitrary direction in the $yz$-plane of the cavity. The angle between the GW direction and the cavity symmetry axis is $\beta$. The external $B$-field is rotated with respect to the cavity symmetry axis by an angle $\alpha$. The GW propagation direction is $\hat{z}'$, and we take the coordinates in the GW frame to be a rotation of the $yz$-plane by $\beta$. The $x$-axis is not shown. 
	}
	\label{fig:orientation_ext_B_field_rot_cavity}
\end{figure}

In the above estimates, we used an approximate value of the coupling coefficient, $\eta_n \sim 0.1$. In this subsection, we demonstrate that this is indeed a typical value for various mode choices in a cylindrical cavity, using the intuition gained from the selection rules of \Sec{currents} to help identify optimally-coupled resonant modes. In a realistic setup, the only aspect of the geometry which we could hope to control is the relative orientation of the cavity axis and the applied magnetic field, which we denote by the angle $\alpha$. The angle of the incoming GW with respect to the cavity axis is denoted by $\beta$. This is shown explicitly in the schematic illustration of the experimental setup in \Fig{orientation_ext_B_field_rot_cavity}. For simplicity, the applied external magnetic field $\B_0$ is taken to be coplanar with the GW wavevector and the cavity symmetry axis $\hat{z}$, but we will also discuss the more general case below. For concreteness, we take cavity dimensions of equal radius and length, $R_{\rm det} = L_{\rm det}$.

The resummed metric is given in \Eq{eq:hfermi} assuming a GW propagating along the $\hat{z}$-direction. In our calculations, we evaluate the effective current $\jveff$ assuming the GW is traveling along the $\hat{z}^\prime$ direction, which we define as a rotation of the $\hat{z}$ axis in the $yz$-plane by an angle $\beta$, as in \Fig{orientation_ext_B_field_rot_cavity}. After determining the effective current in the cavity frame, we numerically evaluate the coupling coefficient $\eta_n$, as in \Eq{eq:formfactor}. Note that $\eta_n$ includes contributions from both azimuthal polarizations of the EM mode (added in quadrature), denoted as the $\pm$ modes mentioned at the beginning of Sec.~\ref{sec:selection_rules} (see Appendices~\ref{app:solenoidal_modes} and \ref{app:energy_density}).

Before explaining the structure of our numerical results in detail, let us consider the coupling of a GW to the cavity mode typically used for axion detection, TM$_{010}$.\footnote{Strictly speaking, this is the TM$_{010-}$ mode in our notation, but for simplicity we leave out the minus index since there is only a single mode function for $m = 0$ rather than a degenerate pair.}  This will enable us to gain some intuition as to why ADMX, HAYSTAC, and other axion experiments only need to re-analyze their existing data to set a limit on GWs, and in doing so why using the full resummed metric in the PD frame is necessary. The electric field component of the TM$_{010}$ mode is $\mathbf{E}_{010} \propto J_{0}( r \, x_{01} / R_{\rm cav}) \, \hat{z}$, where $x_{01}$ is the first root of the Bessel function $J_0(x)$. From the $\E_{010}^* \cdot \jpc$ integrand of the coupling coefficient in \Eq{eq:formfactor}, it is clear that we need only consider the $\hat{z}$ component of the effective current. Taking the orientation of the background magnetic field to be aligned with the cavity axis ($\alpha = 0$, as is typical for axion experiments), and keeping only the first few terms in the long wavelength limit, we find
\begin{align}
     \jcross \cdot \hat{z} \propto \sin{\phi} \,   \sin{\beta} \, \Big(-1 + \frac{2i}{3} \, \wg \,  (z\cos\beta - r\sin\phi\sin\beta) + \order{\wg^2 \, L_\text{det}^2} \Big)
     ~,
\end{align}
where we have switched from Cartesian to cylindrical coordinates. Using the above expression in \Eq{eq:formfactor}, we see that the volume integral of the leading-order contribution to the coupling coefficient $\eta_{010}$ vanishes, owing to the integration of $\sin\phi$ between $0$ and $2\pi$. However, the next-to-leading-order term  contributes to even powers of $\sin\phi$ in the integrand of \Eq{eq:formfactor}, yielding $\eta_{010} > 0$. If we had instead considered  $\jplus$, then we would have obtained a non-zero result only by including even higher order terms. This example clearly shows that keeping the entire series of terms in $\jveff$ (and hence in the metric), rather than only the leading-order term in the expansion, is necessary to see that the TM$_{010}$ mode can be excited by the incoming GW.

\begin{figure}
    \centering
    \includegraphics[width=0.49\textwidth]{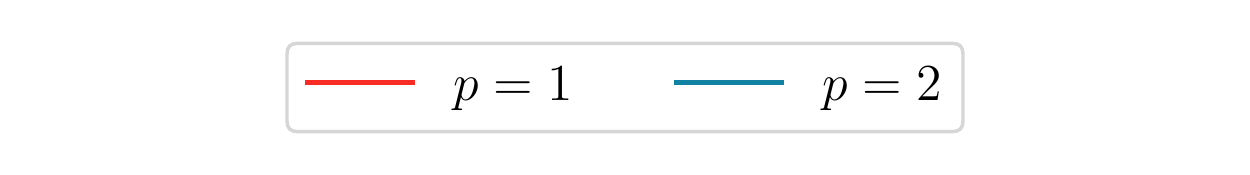}\\
    \includegraphics[width=0.4\textwidth]{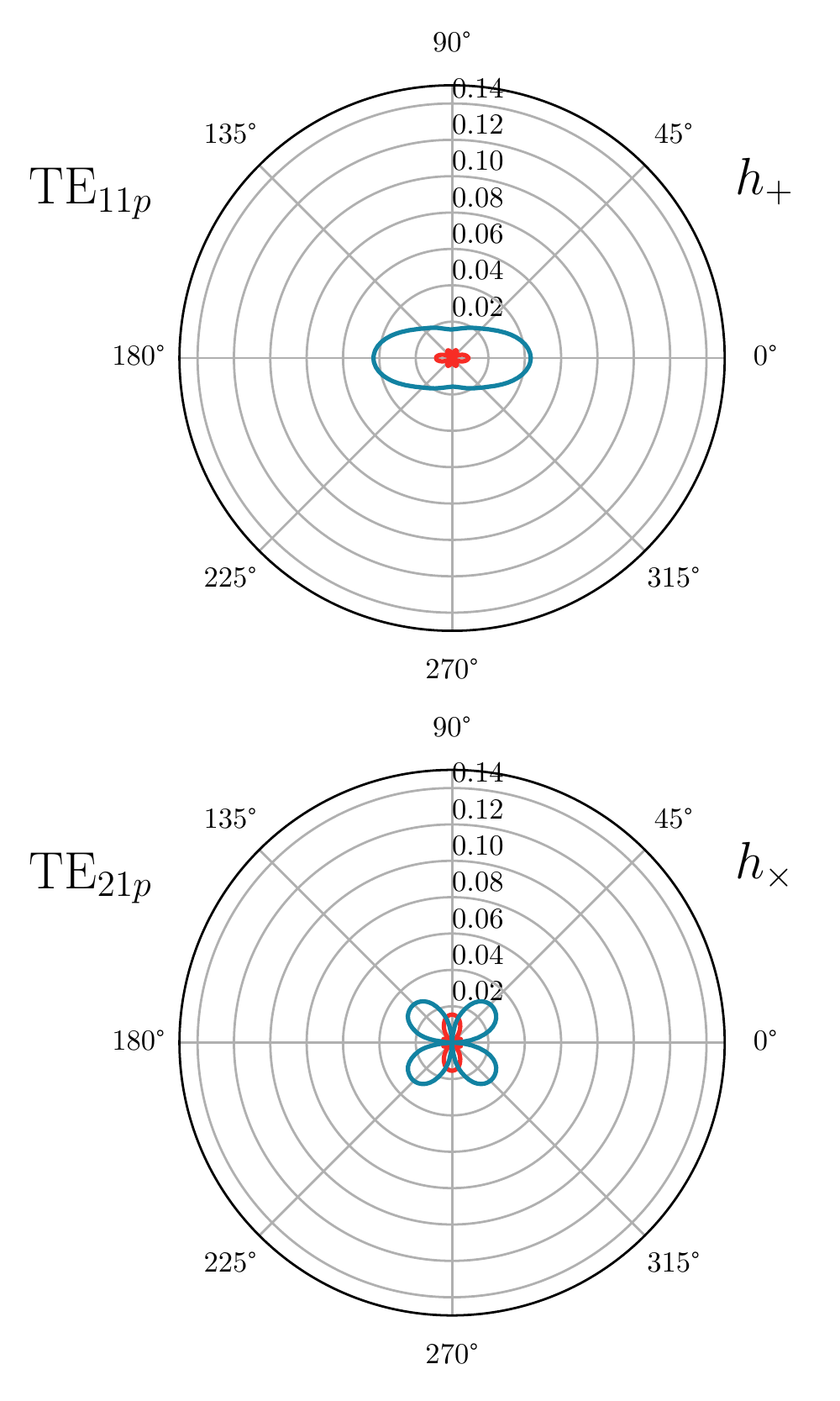}
    \includegraphics[width=0.4\textwidth]{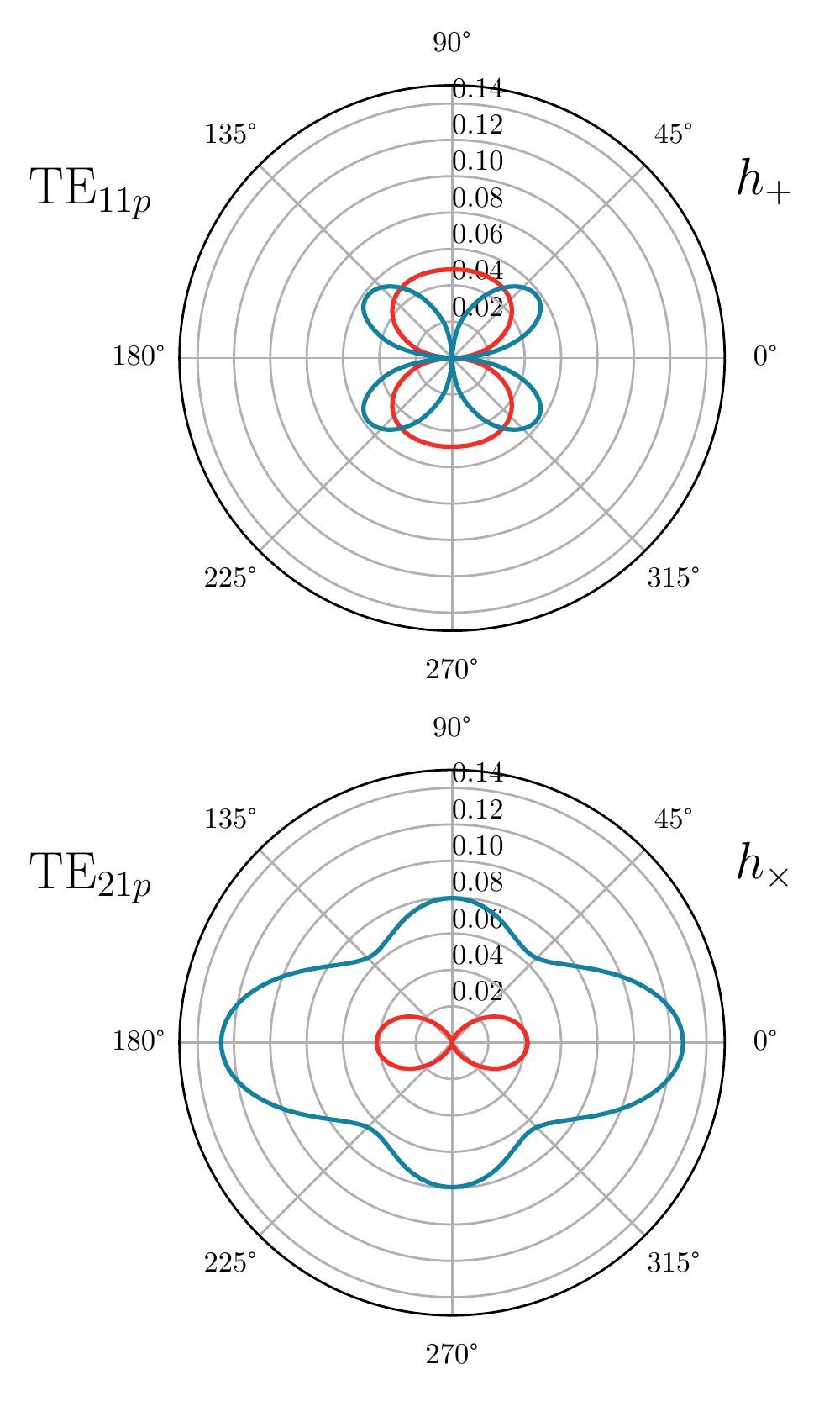}    
    	\caption{The GW-cavity coupling coefficient $\eta_{mnp}$ (the radial component in the polar plane) as a function of the angular direction of the GW propagation ($\beta$) in degrees, for various TE modes and GW polarizations. The notation for the mode index is defined in Eqs.~\eqref{eq:TM_r}--\eqref{eq:TM_z}. Left: the external $B$-field is orthogonal to the cavity symmetry axis ($\alpha=\frac{\pi}{2}$). Right:  the external $B$-field is aligned with the cavity symmetry axis ($\alpha=0$). }
    \label{fig:form_factors_TE}
\end{figure}

\begin{figure}
    \centering
    \includegraphics[width=0.49\textwidth]{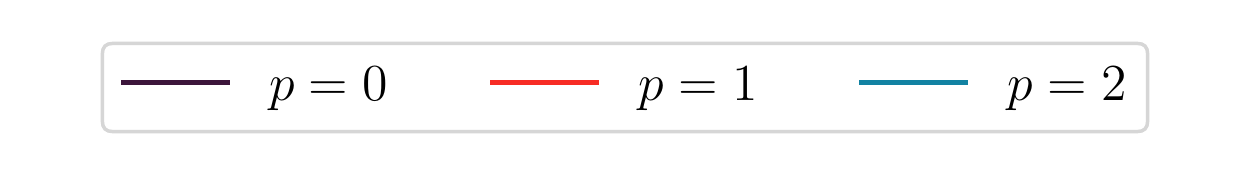}\\
    \includegraphics[width=0.49\textwidth]{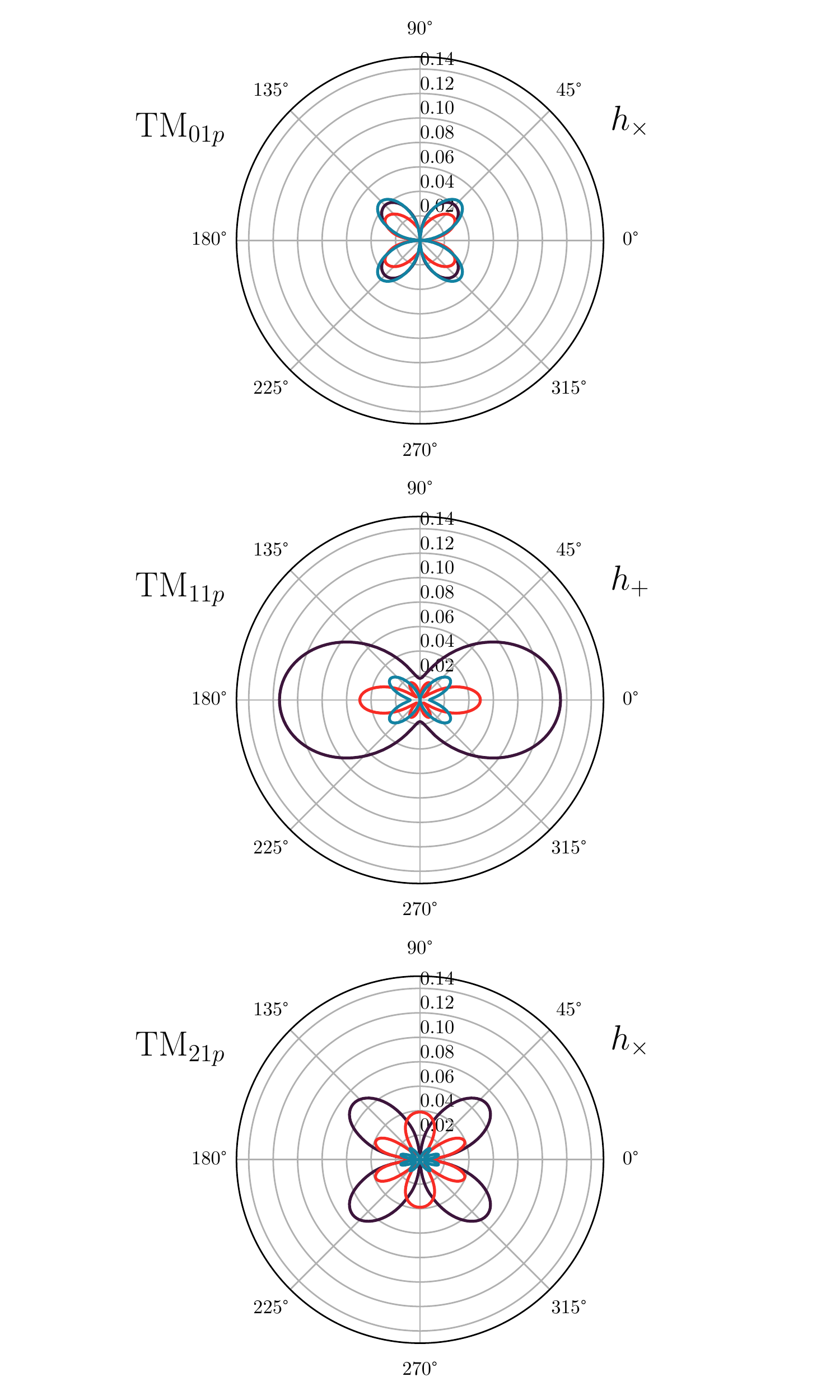}
    \includegraphics[width=0.49\textwidth]{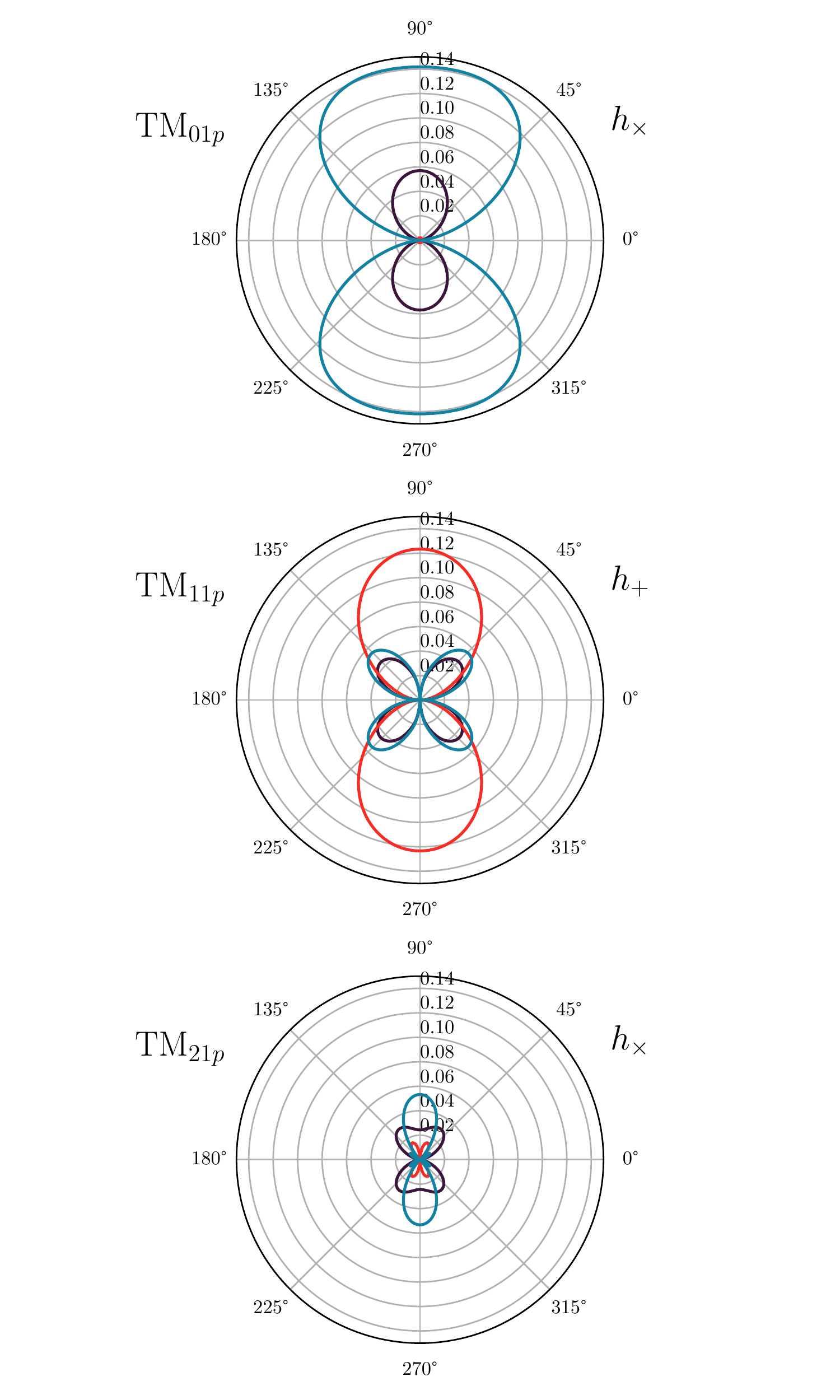}
    	\caption{The GW-cavity coupling coefficient $\eta_{mnp}$ (the radial component in the polar plane) as a function of the angular direction of the GW propagation ($\beta$) in degrees, for various TM modes and GW polarizations. The notation for the mode index is defined in Eqs.~\eqref{eq:TM_r}--\eqref{eq:TM_z}. Left: the external $B$-field is orthogonal to the cavity symmetry axis ($\alpha=\frac{\pi}{2}$). Right:  the external $B$-field is aligned with the cavity symmetry axis ($\alpha=0$). }
    \label{fig:form_factors_TM}
\end{figure}

In Figs.~\ref{fig:form_factors_TE} and \ref{fig:form_factors_TM}, we numerically evaluate the coupling coefficients for various TE and TM modes, respectively. The plots are presented in polar coordinates, for simplicity taking the cavity axis, GW wavevector, and external $B$-field to be coplanar. The radial coordinate is the magnitude of $\eta_n$ and the azimuthal coordinate is the angle $\beta$ of the GW with respect to the cavity axis (see \Fig{orientation_ext_B_field_rot_cavity}). Each row shows a different $m$ mode for a particular choice of GW polarization. In the left (right) columns we take the external $B$-field to be perpendicular (parallel) to the cavity symmetry axis. These results confirm that the typical value of $\eta_n$ for an optimally-coupled mode is $\order{0.1}$; we leave an exhaustive investigation of all possible modes and polarizations to future work.

The coupling coefficient is shown in \Fig{form_factors_TE} for various $\text{TE}_{mnp}$ modes, including the $m = 2$ TE modes that were discussed in \Sec{selection_rules}. In particular, the bottom-right panel of \Fig{form_factors_TE} verifies that this is the only class of modes that couples to a GW when both the external $B$-field and GW wavevector are on-axis with the cavity ($\alpha = \beta = 0$). This selection rule was derived previously from the spin-2 azimuthal dependence of the effective current in \Eq{eq:staticBz}. We also investigate more general geometrical configurations in the remaining panels of Figs.~\ref{fig:form_factors_TE} and \ref{fig:form_factors_TM}, taking the external $B$-field and/or GW wavevector to be unaligned, but still coplanar, with the symmetry axis of the cavity ($\alpha,  \beta \neq 0$). In this case, the lack of cylindrical symmetry implies that a larger set of modes can optimally couple to GWs of various polarizations, while still obeying certain selection rules in the coplanar limit. For instance, the $m = 0$ $\text{TM}_{mnp}$ modes of \Fig{form_factors_TM} only couple to $h_\times$ for the geometries shown. While such selection rules do not generally apply outside of the coplanar limit, these results illustrate the general fact that different modes have different sensitivities to $h_+$ and $h_\times$, potentially enabling the determination of the polarization of a putative GW signal. 

An important point, evident from Figs.~\ref{fig:form_factors_TE} and~\ref{fig:form_factors_TM}, is that different modes have different directional sensitivity. Therefore, if the direction of the GW source is constant over the measurement time, one can in principle spatially localize the GW  source by looking at multiple modes, which may be possible if the signal arises from a compact object merger which is continuously scanning over frequencies (see \Sec{sources}). We leave a detailed examinization of a multi-cavity setup's ability to efficiently characterize GW properties (including the possible role of degeneracies between different polarizations and directions of propagation) to future work.

\subsection{Coupling Coefficient in the Quasistatic Regime}

More generally, we may also consider resonant conversion of GWs to EM fields with a lumped-element LC circuit, where, unlike a cavity, the resonant frequency of the circuit is not tied to the geometrical size of the detector. In this setup, the magnetic field volume can remain large, providing an increased sensitivity for low-frequency signals in the quasistatic limit $\wg \, L_\text{det} \ll 1$, where $L_\text{det}$ is the characteristic size of the experiment.\footnote{Note that in this limit the full time dependence of oscillatory quantities is retained, i.e., $\wg \, t_{\rm int}$ is not a small number, where $t_\text{int}$ is the experimental integration time.} Indeed, this has already been demonstrated as a promising strategy to search for kHz--MHz axions with ABRACADABRA~\cite{Kahn:2016aff,Ouellet:2018beu,Ouellet:2019tlz,Salemi:2021gck}, ADMX-SLIC~\cite{Sikivie:2013laa,Crisosto:2019fcj}, and DM Radio~\cite{Chaudhuri:2014dla,Silva-Feaver:2016qhh}. 

However, we briefly argue here that the GW-EM signal decouples much faster in the quasistatic limit than an axion signal, leading to poor sensitivities at lumped-element experiments. As we showed in Sec.~\ref{sec:selection_rules}, in the low-frequency limit the GW effective current scales as $\jeff \sim \wg^2 \, B_0 \, h \, L_\text{det}$, while as discussed in Sec.~\ref{sec:axion_analogy}, the axion effective current scales as $\jeff \sim \w_a \, B_0 \, \theta_a$. In the limit $\wg \, L_{\rm det} \ll 1$, the GW effective current is therefore suppressed by an additional power of $\wg L_{\rm det}$ compared to an equivalent axion signal with $\theta_a \sim h$. For, e.g., ABRACADABRA-10cm with $L_{\rm det} = 10 \ {\rm cm}$ and $\wg = 10 \ {\rm kHz}$, this amounts to a suppression of $\sim 3 \times 10^{-6}$. That being said, this low-frequency suppression may in fact be a feature for transient signals in quasistatic experiments with broadband capabilities, since the signal strength of a frequency-scanning signal, such as a merger, would grow rapidly even for fixed strain, and could manifest as a chirp-like signal. This parametric suppression motivates considering alternative strategies for GW detection in the kHz--MHz range, including up-conversion of GWs in an oscillating EM background of a driven cavity~\cite{Pegoraro:1978gv,Pegoraro:1977uv, Reece:1984gv, Reece:1982sc, Ballantini:2005am,Bernard:2002ci,Bernard:2001kp, Ballantini:2003nt}, induced by either the GW-EM or mechanical-EM coupling. We will return to these possibilities in future work.

\section{Outlook and Conclusions}
\label{sec:conclusions}

In this work, we have analyzed the interaction between GWs and EM fields. By consistently working in the proper detector frame, which is the reference frame relevant for laboratory experiments, we have shown that axion dark matter haloscope experiments have sensitivity to GHz-scale GWs and only need to reanalyze existing data to set the current best bound on such signals. A generic feature of these setups is that the detailed structure of the signal is strongly tied to the incoming direction and polarization of the GW. Thus, the use of multiple cavities or multiple readout modes of a single cavity may enable the ability to localize the source and determine the polarization of a tentative signal. We have identified sub-solar mass binary mergers and GW emission from superradiant boson clouds as two possible sources in the GHz range to which our setup could theoretically have some sensitivity, though the immediate prospects for detection are not strong. Throughout, we have focused predominantly on EM conversion of GWs in a static background $B$-field using resonant cavity readout, since the signal is parametrically suppressed by $\wg L_{\rm det} \ll 1$ for experiments targeting much lower frequencies. Along these lines, we have noted that most axion experiments designed for lower frequency signals are not favorable for GW detection.

There is, however, at least one important exception to this parametric statement. For instance, using microwave cavities pumped with an oscillating $B$-field considerably improves the low-frequency scaling of the signal. This technique was proposed to detect the mechanical signal induced by GWs through the vibration of the cavity walls  in Refs.~\cite{Caves:1979kq, Pegoraro:1978gv,Pegoraro:1977uv, Reece:1984gv, Reece:1982sc, Ballantini:2005am,Bernard:2002ci,Bernard:2001kp, Ballantini:2003nt}. A similar concept was subsequently applied to the EM signals generated by axion dark matter in Refs.~\cite{Goryachev:2018vjt,Berlin:2019ahk,Berlin:2020vrk,Lasenby:2019prg}. In a future companion paper,  we will extend previous studies to include the direct GW-EM coupling with an oscillating background field, which gives rise to a visible signal also in the idealized case where the cavity is completely isolated from external vibrations. It would also be interesting to compare the generic sensitivity of such a heterodyne detection scheme to the sensitivity of interferometers such as LIGO, extrapolated into the kHz--MHz regime. More generally, we plan to explore the applicability of other axion experimental setups to GW detection, such as those planned for broadband readout at high frequencies~\cite{BREAD:2021tpx}.

A wealth of precious information on the fundamental laws of Nature is encoded in GWs spanning orders of magnitude in frequency and strain. The high-frequency regime, well above known astrophysical sources, is particularly interesting to extract information on early times (and extremely high energies) in the history of the Universe. The interactions of GWs with electromagnetism have long been proposed as a possible avenue towards detection in this frequency regime. In this work, we have further fleshed out details of this approach, leveraging cutting-edge advances in high-$Q$ cavity technology, and we have set the foundations to study future experimental setups that can target this important new frontier in GW detection.

\textit{Acknowledgments.} We thank Masha Baryakhtar, Caterina Braggio, Anna Grassellino, Alex Millar, Jonathan Ouellet, Sam Posen, and Nicolas Yunes for helpful conversations. This  material  is  based  upon  work  supported  by  the U.S.\ Department of Energy,  Office of Science,  National Quantum  Information  Science  Research  Centers,   Superconducting  Quantum  Materials  and  Systems  Center (SQMS) under contract number DE-AC02-07CH11359. 
Fermilab is opertated by the Fermi Research Alliance, LLC under Contract DE-AC02-07CH11359 with the U.S.\ Department of Energy.
The work of SARE was supported in part by SNF Ambizione grant PZ00P2\_193322, \emph{New frontiers from sub-eV to super-TeV}. The work of YK and JSE is supported in part by DOE grant DE-SC0015655. DB is supported by a `Ayuda Beatriz Galindo Senior' from the Spanish `Ministerio de Universidades', grant BG20/00228. IFAE is partially funded by the CERCA program of the Generalitat de Catalunya. The research of DB leading to these results has received funding from the Spanish Ministry of Science and Innovation (PID2020-115845GB-I00/AEI/10.13039/501100011033). DB, RTD, and SARE would like to thank the Galileo Galilei Institute for Theoretical Physics (GGI, Florence), for hospitality during the research program on ``New Physics from The Sky.''

\begin{appendix}

\section{Cavity Mode Functions and Energy Density}
\label{app:cavitymodes}
In general the $E$-fields in a cavity can be expanded in a basis of solenoidal and irrotational cavity modes~\cite{Hill,Collin}
\begin{eqnarray}
	\E(\xv,t)=\sum_n \left \{ e_{sn}(t)\E_{sn}(\xv)+e_{in}(t)\E_{in}(\xv) \right \}.
\end{eqnarray}
The solenoidal mode functions satisfy $\nabla\cdot \E_{sn}=0$ in the cavity volume, while the irrotational mode functions satisfy $\nabla\times\E_{in}=0$. At this point it becomes evident that in general irrotational mode functions are also needed to expand the $E$-field in a cavity, since if we would include only solenoidal mode functions Gauss's law, cf.\ \Eq{eq:InHomo1}, could never be fulfilled. Both solenoidal and irrotational mode functions have vanishing transverse field components on the surface of the cavity. This follows directly from Faraday's law, which as we have argued in \Sec{currents} is not modified by a GW.

In \App{Irrational_modes_cannot_be_enhanced} we show that irrotational modes cannot be resonantly enhanced, hence justifying our anlysis of purely solenoidal modes in the main body. In \App{solenoidal_modes} we give explicit expressions for the two types of solenoidal mode functions, namely TM (transverse magnetic) and TE (transverse electric) modes. In \App{energy_density} we derive the energy density that is stored in a resonantly excited solenoidal cavity mode. 

\subsection{Irrotational modes cannot be resonantly enhanced}\label{app:Irrational_modes_cannot_be_enhanced}

The spatial mode functions fulfill the following relations:
\begin{eqnarray}
	\nabla^2\E_{sn}(\xv)&=&-\w_{sn}^2\E_{sn}(\xv),\\
	\nabla^2\E_{in}(\xv)&=&-\w_{in}^2\E_{in}(\xv),\\
	\int_{\Vcav} dV \E_{sn}(\xv)\cdot\E^*_{sm}(\xv)&=&\delta_{nm}\int_{\Vcav} dV \left|\E_{sn}(\xv)\right|^2,\label{eq:orthogonality_relation}\\
	\int_{\Vcav} dV \E_{in}(\xv)\cdot\E^*_{im}(\xv)&=&\delta_{nm}\int_{\Vcav} dV \left|\E_{in}(\xv)\right|^2,\label{eq:orthogonality_relation_i}\\
	\int_{\Vcav} dV \E_{in}(\xv)\cdot\E^*_{sm}(\xv)&=&0,\label{eq:orthogonality_relation_is}
\end{eqnarray}
where $\w_{sn}$ and $\w_{in}$ are the frequency eigenvalues of the solenoidal and irrotational modes respectively.

For the current that appears in \Eq{eq:vectorizedHelmholtzE} we can assume Ohm's law:
\begin{eqnarray}
	\jv=\sum_n \left \{ \frac{\w_{sn}}{Q_{sn}} e_{sn}(t) \E_{sn}(\xv)+\frac{\w_{si}}{Q_{in}} e_{in}(t) \E_{in}(\xv) \right \},
\end{eqnarray}
where $Q_{sn}$ and $Q_{in}$ are the quality factors for the solenoidal and irrotational modes respectively.

We insert now all expressions into \Eq{eq:vectorizedHelmholtzE} and use that $\nabla\times\nabla\times\E_{sn}=-\nabla^2\E_{sn}+\nabla(\nabla\cdot\E_{sn})=\w_{sn}^2\E_{sn}$ and $\nabla\times\nabla\times\E_{in}=0$:
\begin{eqnarray}
\sum_n e_{sn}\w_{sn}^2\E_{sn}(\xv)+\sum_n \left[\partial_t^2e_{sn}\E_{sn}(\xv)+\partial_t^2e_{in}\E_{in}(\xv)\right]+\sum_n\left[\frac{\w_{sn}}{Q_{sn}}\partial_te_{sn}\E_{sn}(\xv)+\frac{\w_{in}}{Q_{in}}\partial_te_{in}\E_{in}(\xv)\right]=-\partial_t\jveff(\xv,t).\nonumber\\\label{eq:step1}
\end{eqnarray}
Multiplying \Eq{eq:step1} with $\E^*_{in}$, integrating over the cavity volume, and using Eqs.~\eqref{eq:orthogonality_relation} and \eqref{eq:orthogonality_relation_is} gives:
\begin{eqnarray}
	\left(\partial_t^2+\frac{\w_{in}}{Q_{in}}\partial_t\right)e_{in}(t)=\frac{-i\wg\int_{\Vcav}dV \E^*_{in}(\xv)\cdot\jveff(\xv)}{\int_{\Vcav} dV \left|\E_{in}(\xv)\right|^2}e^{i\wg t},\label{eq:eqofmotion_i}
\end{eqnarray}
where we have assumed a monochromatic GW, i.e.\ the effective current density can be written as $\jveff(\xv,t)=e^{i\wg t}\jveff(\xv)$. 
The solution for the irrotational mode functions in \Eq{eq:eqofmotion_i} is given by:
\begin{eqnarray}
	e_{in}(t)=\frac{-i\wg\int_{\Vcav}dV \E^*_{in}(\xv)\cdot\jveff(\xv)}{\int_{\Vcav} dV \left|\E_{in}(\xv)\right|^2}\frac{e^{i\wg t}}{\wg^2-i\frac{\wg\w_{in}}{Q_{in}}}~,\label{eq:sol_irrotational}
\end{eqnarray}
where we have dropped the decaying solutions. From the last factor in Eq.~\eqref{eq:sol_irrotational} we see that on resonance at $\wg = \omega_{in}$ there is no enhancement. This is fundamentally different for the solenoidal time functions. In that case, a similar calculation gives $e_{sn}(t)\sim \frac{1}{\omega^2_{sn}-\wg^2-i\frac{\wg\omega_{sn}}{Q_{sn}}}$, which is resonantly enhanced at $\wg=\omega_{sn}$ for large $Q_{sn}$.

\subsection{Solenoidal mode functions}\label{app:solenoidal_modes}
The solenoidal modes for a cylindrical cavity are classified into TM (transverse magnetic) and TE (transverse electric) modes~\cite{Hill}. The TM$_{mnp}$ modes are:
\begin{eqnarray}
	E_r^\pm&=&- \left[{A_+ \sin m\phi \atop A_-\cos m\phi}\right]\frac{ k_z}{\w_{mnp}^2-k_z^2}\sin(k_zz)J'_{m}(r\gamma_{mn})\gamma_{mn},\label{eq:TM_r}\\
	E_\phi^\pm&=&-\left[{A_+ \cos m\phi \atop -A_-\sin m\phi}\right]\frac{ k_z}{\w_{mnp}^2-k_z^2}\sin(k_zz)\frac{m}{r}J_{m}(r\gamma_{mn}),\label{eq:TM_phi}\\
	E_z^\pm&=& \left[{A_+ \sin m\phi \atop A_- \cos m\phi}\right]J_{m}(\gamma_{mn} r)\cos(k_zz),\label{eq:TM_z}
\end{eqnarray}
where $\omega_{mnp}$ is the resonant frequency of the mode, the $\pm$ at the $E$-fields indicate the upper/lower component in the brackets, $A_\pm$ are constants that are determined by normalizing the modes as in \Eq{eq:orthogonality_relation}, $k_z=\pi p/L_{\rm cav}$, $\gamma_{mn}=x_{mn}/R_{\rm cav}$, $L_{\rm cav}$ and $R_{\rm cav}$ are the length and radius of the cavity respectively and $x_{mn}$ is the $n$th zero of $J_m$. The mode numbers $m$ and $p$ are non-negative integers while $n$ is a positive integer.
The TE$_{mnp}$ modes are:
\begin{eqnarray}
    E_r^\pm&=&\left[{A_+ \cos m\phi \atop -A_- \sin m\phi}\right]\sin \left(k_zz\right)\frac{i\w_{mnp} }{\w_{mnp}^2-k_z^2} \frac{m}{r} J_{m}(r\gamma_{mn}),\label{eq:TE_r}\\
    E_\phi^\pm&=&-\left[{A_+ \sin m\phi \atop A_- \cos m\phi}\right]\frac{i\w_{mnp} }{\w_{mnp}^2-k_z^2}\sin \left(k_zz\right) J_{m}'(\gamma_{mn} r)\gamma_{mn},\label{eq:TE_phi}\\
    E_z^\pm&=&0,\label{eq:TE_z}
\end{eqnarray}
where all variables are defined as in the TM case except that $\gamma_{mn}=x_{mn}'/R_{\rm cav}$, where $x_{mn}'$ is the $n$th zero of $J'_m$. In the main text we denote the mode functions by $\E_{mnp\pm}^{\rm TM}$ or $\E_{mnp\pm}^{\rm TE}$. Often we drop the superscript TM and TE because it is clear from the context what we mean. TM and TE modes are orthogonal to each other in the sense that $\int_{\Vcav} dV\, \E^{\rm TM}\cdot \E^{\rm TE}=0$, and likewise the $+$ and $-$ modes are degenerate and orthogonal for the same $m \neq 0$, $n$, and $p$. Note that an equivalent formalism is to define the modes in a basis of eigenfunctions of the azimuthal rotation generator $\partial_\phi$, which amounts to taking linear combinations of $\E_+$ and $\E_-$; this is convenient in \Sec{selection_rules} when demonstrating the spin-2 nature of the effective current.

\subsection{Energy density}\label{app:energy_density}
In this section we derive the stored energy in a solenoidal cavity mode on resonance. While this is standard material (see, e.g., Ref.~\cite{Hill,jackson_classical_electrodynamics_1999}), we present the derivation here for a self-consistent treatment and also to highlight a pitfall regarding the possible interefence term generated from a strong background field and a weak signal field which has led to some errors in the literature. For better readability we leave out a generic mode index $n$ in all equations in this section.

The energy density $w$ of the EM field in the cavity is given by:
\begin{eqnarray}
w(\xv,t)=\frac{1}{2}\left(\text{Re}\left[\E(\xv,t)\right]^2+\text{Re}\left[\B(\xv,t)\right]^2\right).
\label{eq:energy_density_general}
\end{eqnarray}
We replace the real parts:
\begin{eqnarray}
w(\xv,t)=\frac{1}{8}\left(\left[\E(\xv,t)+\E^*(\xv,t)\right]^2+\left[\B(\xv,t)+\B^*(\xv,t)\right]^2\right),
\end{eqnarray}
and multiply out all terms:
\begin{eqnarray}
w(\xv,t)=\frac{1}{8}\left(\E(\xv,t)^2+2\E(\xv,t)\E^*(\xv,t)+\E^*(\xv,t)^2+\B(\xv,t)^2+2\B(\xv,t)\B^*(\xv,t)+\B^*(\xv,t)^2\right).\label{eq:energy_density_multiplied_out}
\end{eqnarray}
It is important to note that the fields in Eqs.~\eqref{eq:energy_density_general}--\eqref{eq:energy_density_multiplied_out} are the total EM fields, i.e. the $B$-field has a zeroth-order term plus a term of $\mathcal{O}(h)$. In the main text of this paper we assume that it is clear from context when a quantity is first-order. However, to avoid confusion in this section we denote explicitly with $(0)$ and $(1)$ superscripts if a field is zeroth-order or first-order in $h$. We therefore write:
\begin{eqnarray}
\B(\xv,t)&=&\B^{(0)}+\B^{(1)}(\xv,t),\label{eq:$B$-field-parts}\\
\E(\xv,t)&=&\E^{(1)}(\xv,t),\label{eq:$E$-field-parts}
\end{eqnarray}
where $\B^{(1)}(\xv,t)=e^{i\w t}\B^{(1)}(\xv)$, $\E^{(1)}(\xv,t)=e^{i\w t}\E^{(1)}(\xv)$ and $\B^{(0)}=$ const.

The magnetic part of \Eq{eq:energy_density_multiplied_out} is:
\begin{eqnarray}
w_B(\xv,t)&=&\frac{1}{8}\left(\B(\xv,t)^2+2\B(\xv,t)\B^*(\xv,t)+\B^*(\xv,t)^2\right),\\
&=&	
\frac{1}{8}\left(\left(\B^{(0)}\right)^2+2\B^{(0)}\cdot\B^{(1)}(\xv,t)+\left(\B^{(1)}(\xv,t)\right)^2\right.\nonumber\\
& &~~~~~~ + 2 \left(\left|\B^{(0)}\right|^2+\B^{(0)}\left(\B^{(1)}(\xv,t)\right)^*+\left(\B^{(0)}\right)^*\B^{(1)}(\xv,t)+\left|\B^{(1)}(\xv,t)\right|^2\right)\nonumber\\
& &~~~~~~ + \left.\left(\left(\B^{(0)}\right)^*\right)^2+2\left(\B^{(0)}\right)^*\cdot\left(\B^{(1)}(\xv,t)\right)^*+\left(\left(\B^{(1)}(\xv,t)\right)^*\right)^2\right).\label{eq:energy_density_multiplied_out_step1}
\end{eqnarray}
When we perform a time average over all terms in \Eq{eq:energy_density_multiplied_out_step1} we are left with the terms:
\begin{eqnarray}
\bar{w}_B(\xv)=\frac{1}{8}\left(\left(\B^{(0)}\right)^2+2\left(\left|\B^{(0)}\right|^2+\left|\B^{(1)}(\xv)\right|^2\right)+\left(\left(\B^{(0)}\right)^*\right)^2\right).
\end{eqnarray}
The total time-averaged magnetic energy density is:
\begin{eqnarray}
\bar{w}_B(\xv)=\bar{w}_B^{(0)}+\bar{w}_B^{(2)}
\end{eqnarray}
with 
\begin{eqnarray}
\bar{w}_B^{(0)}=\frac{1}{8}\left(\left(\B^{(0)}\right)^2+2\left|\B^{(0)}\right|^2+\left(\left(\B^{(0)}\right)^*\right)^2\right)
\end{eqnarray}
and 
\begin{eqnarray}
\bar{w}_B^{(2)}=\frac{1}{4}\left|\B^{(1)}(\xv)\right|^2.
\end{eqnarray}
Note that there is no first-order term $w_B^{(1)}$ in the energy density because this term vanishes after the time average; note also that in specific experimental realizations like ADMX, which measure only power in specific cavity modes, this term cannot be measured. This fact has lead to some errors in the literature~\cite{PhysRevD.104.023524}.
Similarly, the total time-averaged electric energy density is:
\begin{eqnarray}
	\bar{w}_E(\xv)=\bar{w}_E^{(2)}(\xv)=\frac{1}{4}\left|\E^{(1)}(\xv)\right|^2.
\end{eqnarray}
Therefore the total energy density on top of the energy density $\bar{w}_B^{(0)}$ is:
\begin{eqnarray}
	\bar{w}^{(2)}(\xv)=\bar{w}^{(2)}_B(\xv)+\bar{w}^{(2)}_E(\xv)=\frac{1}{4}\left(\left|\B^{(1)}(\xv)\right|^2+\left|\E^{(1)}(\xv)\right|^2\right)
\end{eqnarray}

The total energy stored in the cavity due to GW excitations is:
\begin{eqnarray}
	W^{(2)}=W^{(2)}_E+W^{(2)}_B=\int_{\Vcav}dV\,\bar{w}^{(2)}(\xv).
\end{eqnarray}
On resonance the magnetic and electric energies are equal:
\begin{eqnarray}
	W_B^{(2)}&=&\frac{1}{4}\int_{\Vcav} dV \,\B^{(1)}(\xv)\cdot \left(\B^{(1)}(\xv)\right)^*\label{eq:WbWe1}\\
	&=&\frac{1}{4}\frac{1}{\w^2}\int_{\Vcav} dV \left(\nabla\times\E^{(1)}\right)\cdot \left(\nabla\times\left(\E^{(1)}\right)^*\right)\label{eq:WbWe2}\\
	&=&\frac{1}{4\w^2}\int_{\Vcav}\left[\left(\E^{(1)}\right)^*\cdot\left(\nabla\times\nabla\times\E^{(1)}\right)+\nabla\cdot\left(\left(\E^{(1)}\right)^{*}\times(\nabla\times\E^{(1)})\right)\right]\label{eq:WbWe3}\\
	&=&\frac{1}{4\w^2}\left[\w^2\int_{\Vcav}dV\left|\E^{(1)}\right|^2+\int_{S}d\bm{S}\cdot\left(\left(\E^{(1)}\right)^*\times(\nabla\times\E^{(1)})\right)\right]\label{eq:WbWe4}\\
	&=&\frac{1}{4}\int_{\Vcav}dV\left|\E^{(1)}\right|^2=W_E^{(2)}.\label{eq:WbWe5}
\end{eqnarray}
When going from \Eq{eq:WbWe1} to \Eq{eq:WbWe2} we have used that the $B$-field fulfills Maxwell's equations and we have assumed that we are on resonance, where $\w \equiv \wg$ is the resonance frequency of the solenoidal cavity mode that we consider. In the second step (\Eq{eq:WbWe2} to \Eq{eq:WbWe3}) we have used that
\begin{eqnarray}
	\left(\nabla\times \E^{(1)}\right)\cdot\left(\nabla\times\left(\E^{(1)}\right)^{*}\right)=\epsilon_{ijk}\partial_jE_k^{(1)}\epsilon_{ilm}\partial_l\left(E^{(1)}_m\right)^*\nonumber\\ \overset{\rm PI}{=}-\epsilon_{ijk}\epsilon_{ilm} \left(E_m^{(1)}\right)^*\partial_l\partial_j E^{(1)}_k+\epsilon_{ijk}\epsilon_{ilm}\partial_l\left(\left(E^{(1)}_m\right)^*\partial_jE^{(1)}_k\right)\nonumber\\=\left(\E^{(1)}\right)^*\cdot\left(\nabla\times\nabla\times\E^{(1)}\right)+\nabla\cdot\left(\left(\E^{(1)}\right)^*\times(\nabla\times\E^{(1)})\right)\nonumber
\end{eqnarray}
where we have used a partial integration (PI) by going from the first to the second equation.
Finally when going from \Eq{eq:WbWe3} to \Eq{eq:WbWe4} we used $\nabla\times\nabla\times\E^{(1)}=\w^2\E^{(1)}$ for a solenoidal mode; here $S$ is the surface of the cavity. Furthermore we used Gauss's theorem. In the last step we used that on the surface of a perfect conductor $d\bm{S}$ is parallel to $\E^{(1)}$ and therefore the last term vanishes.

To summarize, the total energy that is stored in the cavity and due to the interaction with GWs is:
\begin{eqnarray}
	W^{(2)}=\frac{1}{2}\int_{\Vcav}\,dV\,\left|\E^{(1)}(\xv)\right|^2. \label{eq:energy_density_general_expression}
\end{eqnarray}
In the case that $\E^{(1)}$ is the sum of two $\E_{\pm}$ modes, cf. \Eq{eq:TM_r} to \Eq{eq:TE_z}, $W^{(2)}$ is the squared sum of both contributions since $\E_\pm$ modes are orthogonal: $W^{(2)}=\frac{1}{2}\int_{\Vcav}\,dV\,\left|\E^{(1)}_{+}(\xv)\right|^2+\frac{1}{2}\int_{\Vcav}\,dV\,\left|\E^{(1)}_{-}(\xv)\right|^2$.

\end{appendix}

\bibliography{bibliography}

\end{document}